\documentclass[pdflatex,sn-nature]{sn-jnl}

\usepackage{graphicx}
\usepackage{amsmath,amssymb,amsfonts}
\usepackage{amsthm}
\usepackage{booktabs}
\usepackage{multirow}
\usepackage{xcolor}
\usepackage{textcomp}
\usepackage{manyfoot}
\usepackage{algorithm}
\usepackage{algorithmic}

\theoremstyle{thmstyleone}

\theoremstyle{thmstyletwo}

\theoremstyle{thmstylethree}

\unnumbered

\begin{document}

\title{Event-driven signal reconstruction through neuromorphic compressive sensing}

\author*[1]{\fnm{Zeru} \sur{Fang}}
\email{z.fang25@imperial.ac.uk}

\author[1]{\fnm{Yanzhen} \sur{Liu}}

\author[1]{\fnm{Geoffrey Ye} \sur{Li}}

\affil[1]{%
  \orgdiv{Department of Electrical and Electronic Engineering},
  \orgname{Imperial College London},
  \orgaddress{%
    \city{London},
    \postcode{SW7 2AZ},
    \country{United Kingdom}%
  }%
}

\abstract{
Compressive sensing (CS) exploits intrinsic signal sparsity for efficient representation of high-dimensional data. However, conventional CS relies on fixed-length measurement representations and dense reconstruction operations, leaving communication and computational costs largely determined by system dimensions rather than signal sparsity. Here, we propose a neuromorphic CS framework centred on a spike-driven learned iterative shrinkage-thresholding algorithm (S-LISTA). By representing compressed measurements and reconstruction updates as sparse events, the framework links intrinsic signal sparsity to the spatiotemporal sparsity of spiking neural networks (SNNs), extending the benefits of sparsity across sensing, transmission and reconstruction. Communication and computational costs consequently depend on event activity, allowing sparse representations to translate into resource savings. Extensive experiments demonstrate substantial reductions in transmitted data volume and estimated reconstruction energy while maintaining competitive reconstruction and downstream task performance. The framework also exhibits robustness under challenging channel conditions. These findings establish neuromorphic CS as a promising paradigm for signal reconstruction in resource-constrained scenarios.

}


\maketitle

\section{Introduction}
\label{sec:introduction}

Compressive sensing (CS) maps a high-dimensional signal to fewer linear measurements and exploits its sparsity in a transform domain to recover the source~\cite{1614066,4472240}. It is used in magnetic resonance imaging~\cite{10578304}, electrocardiogram (ECG) compression~\cite{fira2022study}, and wireless sensor network data collection~\cite{6384860}. Moreover, the resulting low-dimensional measurements provide a compact representation for data transmission, making CS a useful tool for data compression~\cite{6384860}. In Internet of Things (IoT) applications, connected devices generate and exchange large volumes of data under constraints on bandwidth, computing capacity, memory, and energy~\cite{8016573,shi2016edge}. Reducing transmitted data therefore makes CS attractive for these resource-constrained applications. However, conventional CS typically represents measurements as dense, fixed-length vectors, so communication costs depend on measurement dimensionality without further benefiting from sparsity in the signal. Source reconstruction also commonly relies on dense, continuous-valued matrix operations, incurring high computational and energy costs~\cite{gregor2010learning,horowitz20141}. These costs limit the use of conventional CS on resource-constrained edge devices~\cite{8016573} and motivate extending the use of sparsity beyond source representation to measurement representation, transmission, and reconstruction.

Biological nervous systems provide a model for information processing based on sparse neural activity. Sensory pathways encode stimuli into temporally structured spike trains~\cite{dettner2016temporal}. Neurons integrate synaptic inputs and emit discrete spikes when their membrane potentials cross a firing threshold~\cite{stuart2015dendritic,gerstner2002spiking}. Spiking neural networks (SNNs) abstract these principles into computational models that represent and process information through discrete events rather than dense, continuous activations~\cite{MAASS19971659,roy2019towards}. Event-driven operation limits synaptic computation to spike events and can reduce computational cost when neural activity is sparse~\cite{doi:10.1126/science.1254642,davies2018loihi}. These properties support the use of SNNs in visual and auditory classification, object detection, semantic segmentation, and event-based video reconstruction~\cite{amir2017low,cramer2020heidelberg,kim2020spiking,kim2022beyond,zhu2022event}. The event-driven operation of SNNs allows the signal sparsity exploited by CS to be reflected directly in the computation. Sparse signals can therefore be processed through sparse computation, further reducing resource consumption. This correspondence between signal sparsity and computational sparsity provides a promising foundation for resource-efficient CS.

Previous studies have shown that SNNs can perform sparse coding for signal reconstruction~\cite{woods2018fast,davies2018loihi}. Spiking implementations of the locally competitive algorithm (LCA) are a prominent example. They balance reconstruction fidelity against representation sparsity through recurrent neural dynamics that evolve towards convergence~\cite{rozell2008sparse,woods2018fast,davies2018loihi}. Consequently, representation quality depends on the available inference time. With a fixed number of updates, the recovery process may terminate before reaching an accurate solution. Algorithm unfolding addresses this constraint by mapping a prescribed number of optimization iterations onto a fixed-depth trainable architecture~\cite{8253590,9363511}. The learned iterative shrinkage-thresholding algorithm (LISTA) unfolds the iterative shrinkage-thresholding algorithm (ISTA). It retains the residual-correction and thresholding steps while learning the associated operators and thresholds from data~\cite{daubechies2004iterative,gregor2010learning,chen2018theoretical}. The network depth therefore fixes the number of recovery updates without requiring recurrent dynamics to converge. However, the residual correction in each LISTA layer still relies on dense, continuous-valued matrix operations. Repeating these operations across layers requires many multiply-accumulate (MAC) operations and incurs substantial computational cost. The need to reduce these computational and energy costs motivates a neuromorphic unfolding method that combines fixed-depth recovery with sparse event-driven computation.

Here, we introduce a neuromorphic compressive sensing framework centred on a spike-driven learned iterative shrinkage-thresholding algorithm (S-LISTA). At the transmitter, an SNN projects source observations into lower-dimensional measurements and represents them as binary spike events for transmission. At the receiver, S-LISTA realizes the unfolded LISTA recovery process through spiking dynamics. Across a fixed number of unfolded layers, S-LISTA forms the sparse code by accumulating discrete spike updates. The residual membrane potential then supplies a continuous correction on the selected support before the final readout. The sensing and recovery components are trained end to end together with a learnable dictionary, while task supervision acts directly on the sparse code to encourage the retention of information relevant to downstream inference~\cite{mairal2012task,zhang2010discriminative,jiang2013label}. Under the address-event representation (AER) transmission protocol, the proposed framework reduces transmitted bits by 96.1\% on DVS-Gesture (DVS-G) and 96.5\% on the Spiking Heidelberg Digits (SHD) relative to the artificial neural network (ANN) baselines with 8-bit quantized measurements. At matched unfolded depth, its event-driven reconstruction requires only $1/22.8$ and $1/82.4$ of the estimated energy consumed by ANN LISTA on the two datasets, respectively. Experiments also demonstrate robustness in reconstruction and downstream task performance under challenging channel conditions, including fading not encountered during training. Overall, this work connects signal sparsity with the spatiotemporal sparsity of SNNs, offering a promising paradigm for the design of future resource-constrained edge systems.

\section{Results}
\label{sec:results}

\subsection{Neuromorphic compressive sensing architecture}
\label{subsec:framework}

\begin{figure*}[hbpt]
    \centering
    \includegraphics[width=\textwidth]{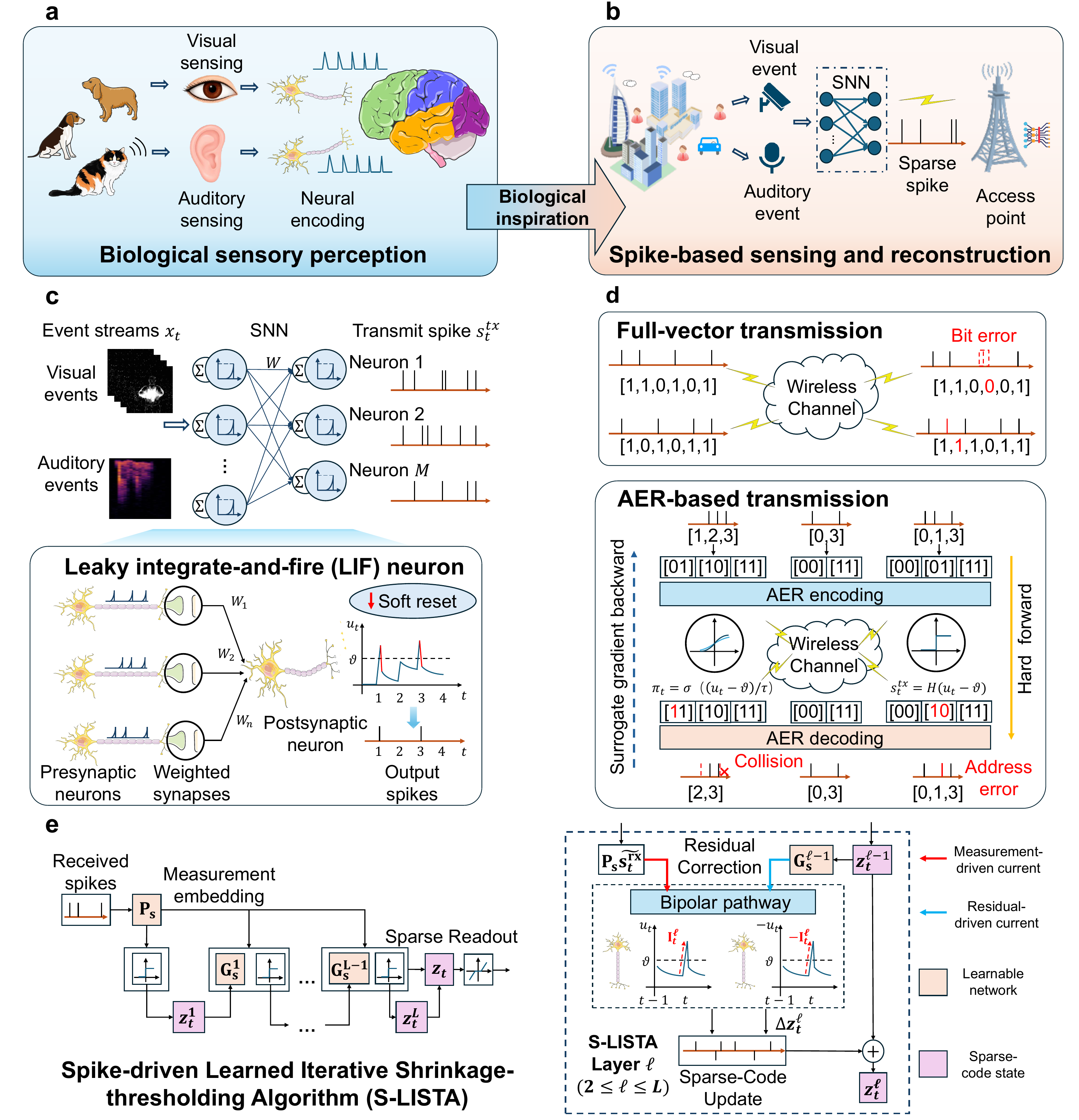}
    \caption{\textbf{Neuromorphic compressive sensing with S-LISTA.}
\textbf{a}, Biological inspiration from sensory spike encoding.
\textbf{b}, Framework overview of spike-based measurement, wireless transmission, and reconstruction.
\textbf{c}, Spiking measurement network and leaky integrate-and-fire (LIF) neuron dynamics.
\textbf{d}, Full-vector and address-event representation (AER) transmission over a wireless channel.
\textbf{e}, Illustration of the S-LISTA architecture. Medical illustrations provided by Servier Medical Art, licensed under CC BY 4.0.}
    \label{fig:framework}
\end{figure*}

Figure~\ref{fig:framework}a illustrates how biological sensory systems encode visual and auditory information into sparse spikes for neural transmission and processing. Inspired by this principle, we propose the neuromorphic compressive sensing framework shown in Fig.~\ref{fig:framework}b, which uses spikes for measurement, wireless transmission, and reconstruction. The framework operates as follows.

At the transmitter, sensors on edge devices collect large volumes of data that require compression before transmission. An SNN measurement network uses leaky integrate-and-fire (LIF) neurons to map these data into low-dimensional measurement spikes, as shown in Fig.~\ref{fig:framework}c. The resulting spikes are transmitted over a wireless channel using either full-vector transmission or AER, as illustrated in Fig.~\ref{fig:framework}d. Full-vector transmission represents all measurement coordinates, whereas AER encodes only the addresses of active spikes~\cite{842110}. When spike activity is sufficiently sparse, the savings from omitting inactive coordinates outweigh the address-encoding overhead, reducing the transmitted data volume.

The receiver reconstructs the original data from the received measurement spikes while preserving information for potential downstream tasks. To exploit sparse computation for energy-efficient reconstruction, we deploy the proposed S-LISTA at the receiver, as shown in Fig.~\ref{fig:framework}e. S-LISTA preserves the LISTA update structure and iteratively updates a sparse code by accumulating positive and negative spikes. At the final readout, the residual membrane potential provides a continuous correction to the activated coefficients. A learned dictionary then reconstructs the original data from the resulting sparse code. The detailed S-LISTA formulation is provided in Methods. We use joint training to optimize the measurement network, reconstruction network, and dictionary, with task supervision on the sparse code to retain information relevant to downstream tasks.

\begin{figure*}[htbp]
    \centering
    \includegraphics[width=\textwidth]{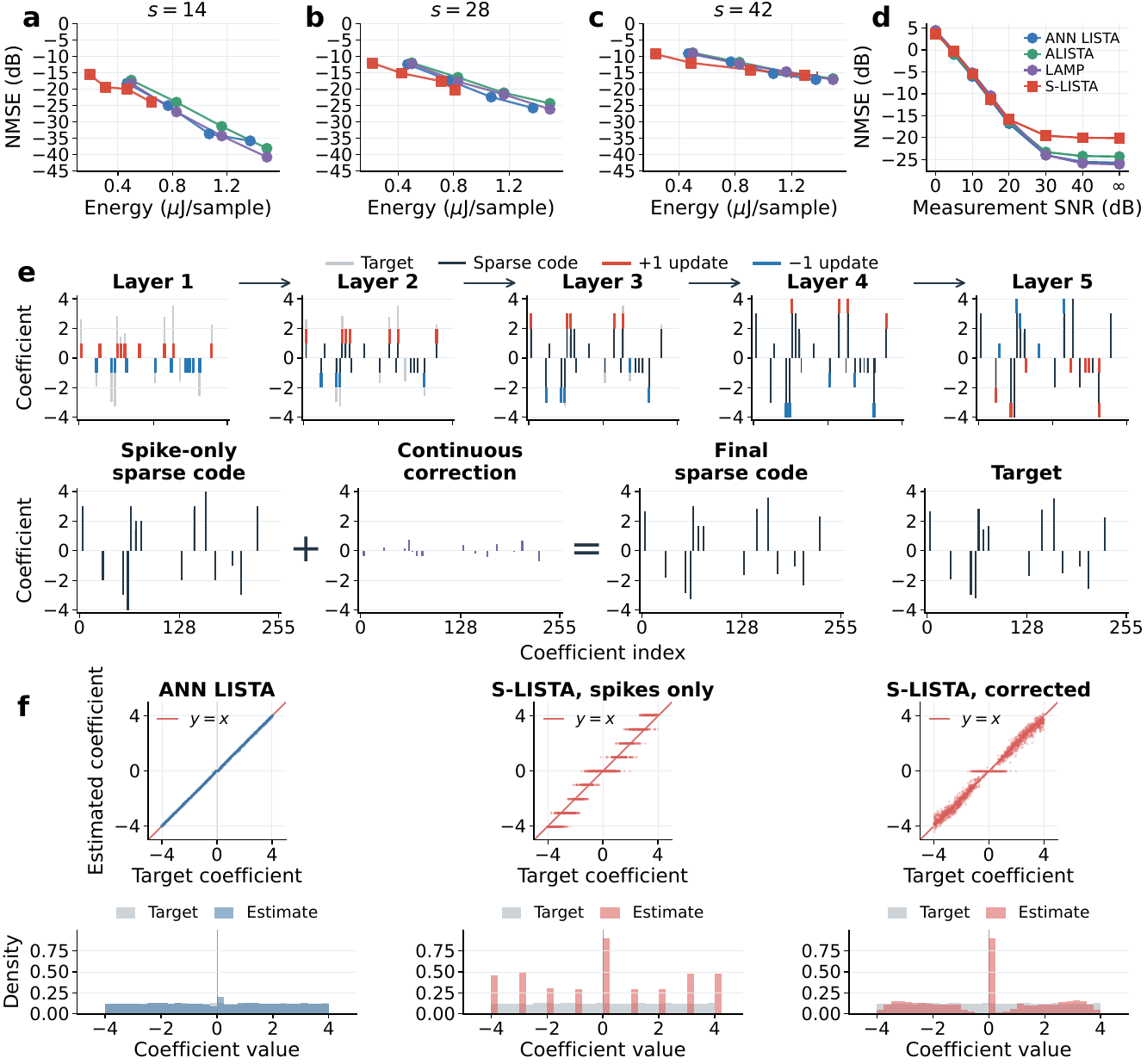}
\caption{\textbf{Sparse reconstruction on synthetic signals.}
\textbf{a--c}, Normalized mean squared error (NMSE) versus estimated reconstruction energy for $s=14$, $28$, and $42$, respectively, where $s$ is the number of nonzero coefficients.
\textbf{d}, NMSE versus measurement signal-to-noise ratio (SNR) at $s=28$; $\infty$ denotes noise-free measurements.
\textbf{e}, Layerwise spike updates and continuous correction for $s=14$, five unfolded layers, and one time step.
Red and blue segments denote $+1$ and $-1$ updates; dark and grey stems show the accumulated sparse code and target, respectively.
The lower row compares the spike-only code, continuous correction, final code, and target.
\textbf{f}, Reconstructed versus target coefficients and their distributions across 2,000 signals, restricted to nonzero target coordinates.}
    \label{fig:toy}
\end{figure*}

\subsection{S-LISTA enables energy-efficient sparse reconstruction}
\label{subsec:toy_results}

We first evaluate S-LISTA on synthetic sparse signals. Each signal has a dimension of $N=256$, with exactly $s$ nonzero coefficients drawn uniformly from $[-4,4]$. Measurements are obtained using a fixed random Gaussian matrix with $M=141$, corresponding to a measurement ratio of $55.1\%$. An identity dictionary is used for reconstruction. We compare S-LISTA with ANN LISTA, analytic LISTA (ALISTA)~\cite{liu2019alista}, and learned approximate message passing (LAMP)~\cite{borgerding2017amp}. Results are averaged over five random seeds. Increasing the unfolded depth reduces S-LISTA's reconstruction error with a moderate increase in estimated energy, as shown in Figs.~\ref{fig:toy}a--c. At $s=42$, the improvement in NMSE becomes smaller as the unfolded depth increases to 20. ANN LISTA and LAMP achieve lower NMSE when more computation is available. At comparable NMSE, S-LISTA reduces estimated energy consumption by up to $57.3\%$ relative to the ANN baselines. These results support the use of S-LISTA for sparse recovery under a limited energy budget.

We then evaluate noise robustness at $s=28$ by adding Gaussian noise to the compressed measurements, as shown in Fig.~\ref{fig:toy}d. Relative to the noiseless condition, S-LISTA's NMSE increases by $0.54$ and $4.24$~dB at measurement signal-to-noise ratios (SNRs) of $30$ and $20$~dB, respectively. At SNRs of $20$~dB and below, S-LISTA achieves similar reconstruction performance to the ANN baselines.

Continuous correction improves the precision of the sparse code obtained through discrete spike accumulation. In the example in Fig.~\ref{fig:toy}e, it reduces NMSE from $-15.14$ to $-26.89$~dB. The coefficient comparisons in Fig.~\ref{fig:toy}f further show that continuous correction reduces the discretization error of spike-only accumulation. However, some small-magnitude coefficients remain unrecovered because the corresponding membrane potentials do not reach the firing threshold. This thresholding behaviour can alter the sparse-code distribution, motivating a learnable dictionary adapted to the codes produced by S-LISTA.

\begin{figure*}[htbp]
\centering
\includegraphics[width=\textwidth]{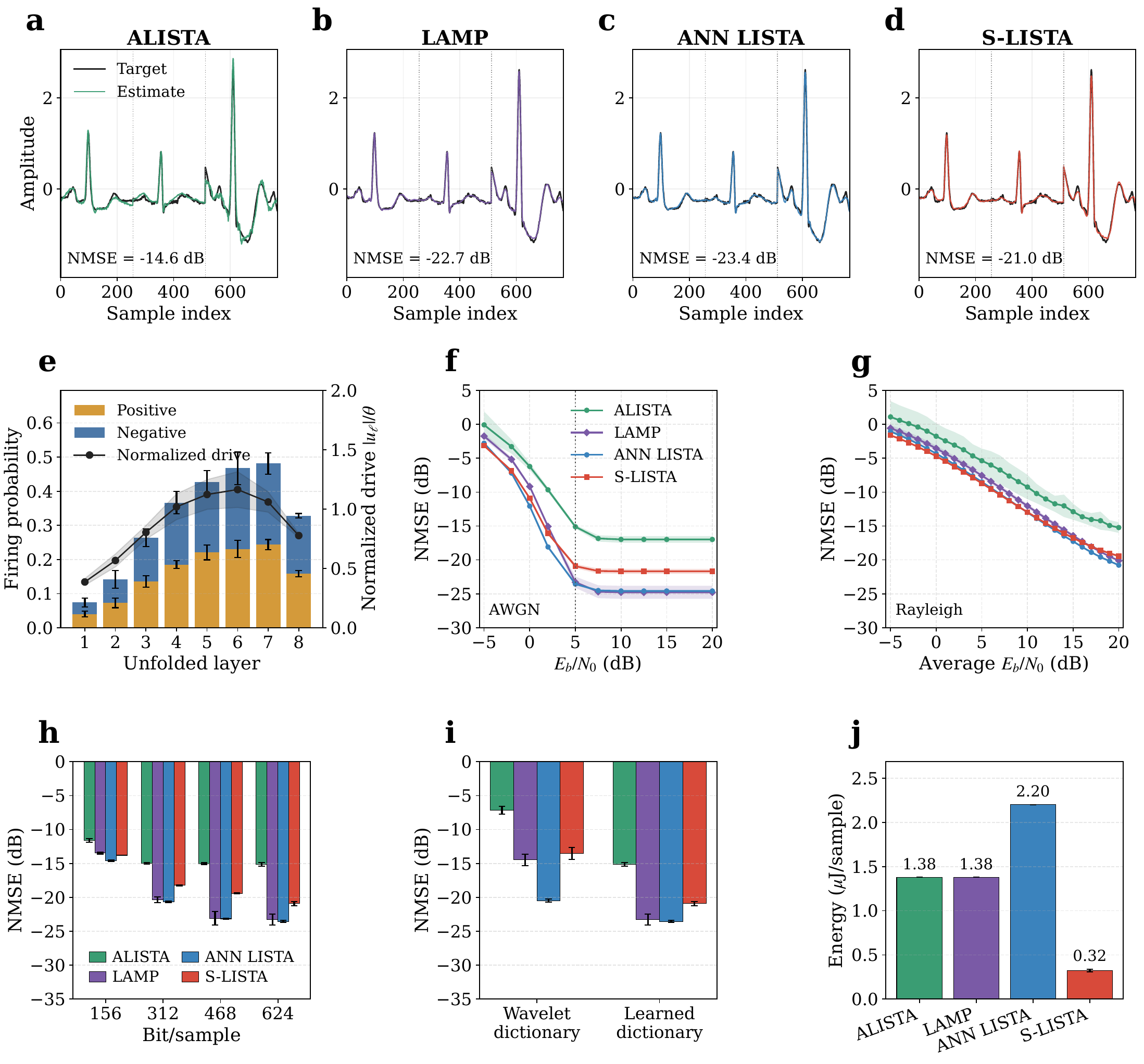}
\caption{\textbf{ECG reconstruction with a learnable dictionary.}
\textbf{a--d}, Example reconstructions using ALISTA, LAMP, ANN LISTA, and S-LISTA with learned sensing matrices and dictionaries and 8-bit measurements.
\textbf{e}, Layerwise positive and negative firing probabilities and mean absolute drive normalized by the firing threshold.
\textbf{f,g}, NMSE over additive white Gaussian noise (AWGN) and quasi-static Rayleigh fading, respectively. Models are trained over AWGN at 5~dB, marked by the dotted line in \textbf{f}.
\textbf{h}, NMSE versus transmitted bits per sample with learned sensing matrices and dictionaries.
\textbf{i}, NMSE using fixed Gaussian sensing and a Symlet-4 dictionary versus jointly learned sensing matrices and dictionaries.
\textbf{j}, Estimated reconstruction energy, excluding dictionary readout.
Panels \textbf{a--d,h,i} use AWGN at 5~dB.
Panels \textbf{a--d} show individual reconstruction examples rather than averages across seeds.
Results in \textbf{e--j} are averaged over five random seeds; error bars and shaded bands indicate one standard deviation.}
\label{fig:ecg}
\end{figure*}

\subsection{Learnable sensing and dictionaries improve ECG reconstruction}
\label{subsec:ecg_results}

We next evaluate S-LISTA for ECG reconstruction using the MIT-BIH Arrhythmia Database. Each heartbeat contains $N=256$ samples and is compressed into $M=78$ measurements. To adapt reconstruction to the sparse codes produced by S-LISTA, we compare a jointly learned sensing matrix and dictionary with a fixed Gaussian sensing matrix and a Symlet-4 wavelet dictionary.

The example reconstructions in Figs.~\ref{fig:ecg}a--d preserve the principal morphology of the cardiac waveforms. With learned sensing matrices and dictionaries, S-LISTA achieves a mean NMSE of $-20.92$~dB, outperforming ALISTA by 5.81~dB but trailing LAMP and ANN LISTA by 2.37 and 2.64~dB, respectively. Its reconstruction updates remain sparse: $68.13\%$ of coefficient-layer positions generate no spike event, as shown in Fig.~\ref{fig:ecg}e.

S-LISTA becomes more competitive under challenging channel conditions. At an SNR of 2~dB under additive white Gaussian noise (AWGN), S-LISTA achieves a lower NMSE than LAMP, as shown in Fig.~\ref{fig:ecg}f. When the models trained over AWGN are evaluated under unseen Rayleigh fading, S-LISTA achieves the lowest NMSE among the four methods at all tested SNRs from $-5$ to 9~dB, as shown in Fig.~\ref{fig:ecg}g. S-LISTA also maintains competitive reconstruction performance under a limited transmission budget. At 156 bits per sample, it achieves an NMSE of $-13.79$~dB, outperforming LAMP by 0.34~dB and trailing ANN LISTA by 0.83~dB in Fig.~\ref{fig:ecg}h. Increasing the transmission budget improves all methods, although S-LISTA benefits less than ANN LISTA and LAMP.

Joint training of the sensing matrix and dictionary reduces NMSE by 3.07--8.82~dB across the four methods, as shown in Fig.~\ref{fig:ecg}i. For S-LISTA, NMSE decreases from $-13.56$ to $-20.92$~dB, an improvement of 7.37~dB. Figure~\ref{fig:ecg}j shows that its estimated reconstruction energy is $0.322\,\mu\mathrm{J}$ per sample, compared with $1.378\,\mu\mathrm{J}$ for LAMP and $2.202\,\mu\mathrm{J}$ for ANN LISTA, excluding the final dictionary readout. The corresponding energy reductions are $76.64\%$ and $85.39\%$. Together, these results show that joint learning improves ECG reconstruction while preserving sparse computation, with competitive performance under limited transmission budgets and challenging channel conditions.

\begin{figure*}[htbp]
\centering
\includegraphics[width=0.9\textwidth]{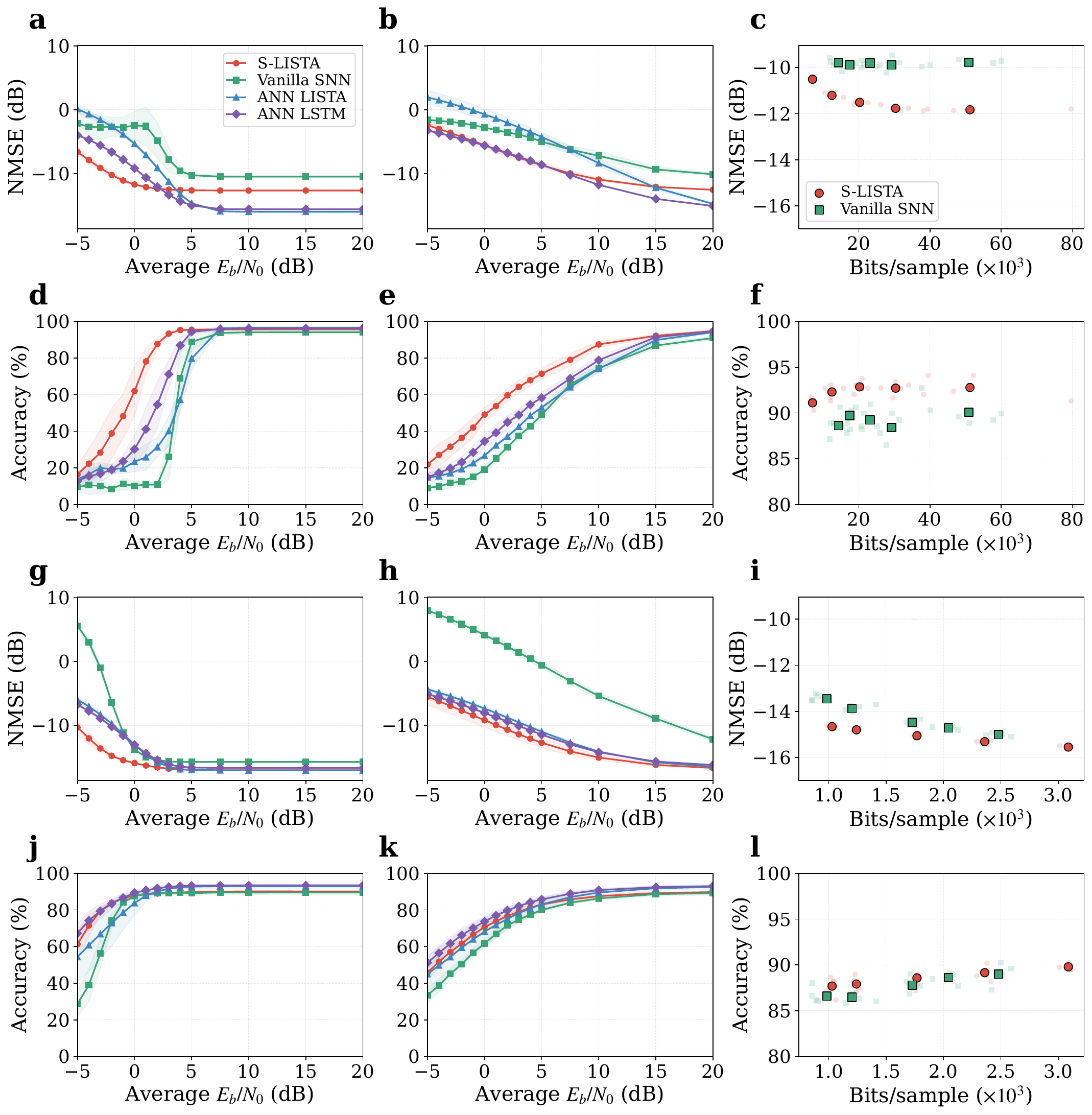}
\caption{\textbf{Reconstruction, downstream task performance, and transmission cost on neuromorphic datasets.}
\textbf{a,b}, Reconstruction NMSE on DVS-G over AWGN and Rayleigh fading, respectively.
\textbf{d,e}, Corresponding downstream task accuracy.
\textbf{c,f}, NMSE and downstream task accuracy versus AER bits per sample on DVS-G under Rayleigh fading at 20~dB.
\textbf{g,h,j,k}, Corresponding SNR scans on SHD.
\textbf{i,l}, Corresponding AER transmission results on SHD.
SNR scans use full-vector transmission.
All models are trained over AWGN at 10~dB and evaluated without retraining.
Curves and large markers show means over five independent seeds; shaded bands indicate standard deviations, and faint markers show individual-seed results.}
\label{fig:event_performance}
\end{figure*}

\subsection{Neuromorphic compressive sensing reduces transmission cost on event data}

We further evaluate the complete framework on DVS-G and SHD as representative visual and auditory event datasets. We compare reconstruction and downstream task performance using S-LISTA, a vanilla SNN reconstructor, and ANN implementations of LISTA and long short-term memory (LSTM) networks. The ANN baselines transmit continuous measurements quantized to 8 bits, whereas the SNN methods transmit binary measurement spikes. Table~\ref{tab:main_event_results} summarizes performance under Rayleigh fading at 20~dB. On both datasets, S-LISTA transmits $87.5\%$ fewer bits than the ANN baselines. On DVS-G, its reconstruction performance is lower than that of the ANN methods, but its downstream task performance remains comparable. It also outperforms the vanilla SNN in both reconstruction and downstream task accuracy. On SHD, S-LISTA achieves an NMSE of $-16.66$~dB, comparable to $-16.45$~dB for ANN LISTA, while its downstream task accuracy is approximately 3 percentage points lower.

S-LISTA shows a clear downstream task advantage under challenging channel conditions on DVS-G. At 2~dB over AWGN, it maintains an accuracy of $87.71\%$, compared with $31.32\%$ for ANN LISTA, as shown in Fig.~\ref{fig:event_performance}d. Under unseen Rayleigh fading, S-LISTA achieves the highest mean downstream task accuracy at every tested SNR from $-5$ to 20~dB in Fig.~\ref{fig:event_performance}e. These results show that S-LISTA better preserves task-relevant information under challenging channel conditions.

On SHD, S-LISTA achieves reconstruction performance comparable to the ANN methods and shows greater robustness to unseen fading than the vanilla SNN. Under Rayleigh fading, it achieves the lowest mean NMSE among all methods at every tested SNR from $-5$ to 20~dB in Fig.~\ref{fig:event_performance}h. Its NMSE is lower than that of the vanilla SNN throughout the tested range, with a minimum gap of approximately 4.49~dB. Its mean downstream task accuracy is also higher than that of the vanilla SNN at every tested SNR in Fig.~\ref{fig:event_performance}k.

Under AER transmission, S-LISTA provides better reconstruction than the vanilla SNN at comparable transmission budgets, as shown in Figs.~\ref{fig:event_performance}c,f,i,l. On DVS-G, this reconstruction advantage is accompanied by higher downstream task accuracy across the tested operating points. At approximately $30\times10^3$ bits per sample, S-LISTA achieves an NMSE improvement of 1.87~dB and an accuracy increase of 4.31 percentage points relative to the vanilla SNN. On SHD, S-LISTA also achieves lower reconstruction error while maintaining similar downstream task performance. At approximately $1.7\times10^3$ bits per sample, its NMSE is 0.58~dB lower, with both methods achieving approximately $88\%$ accuracy. These results show that S-LISTA makes more effective use of the transmission budget for reconstruction while preserving downstream task performance.

\begin{table*}[t]
\centering
\caption{\textbf{Main reconstruction and downstream task results under Rayleigh fading at 20~dB.}
Results use full-vector transmission and are reported as means $\pm$ standard deviations over five independent seeds.}
\label{tab:main_event_results}
\begingroup
\footnotesize
\setlength{\tabcolsep}{5pt}
\renewcommand{\arraystretch}{1.08}
\begin{tabular*}{\textwidth}{@{\extracolsep{\fill}}llccc@{}}
\toprule
Dataset & Method & NMSE (dB) & Accuracy (\%) & Bits/sample \\
\midrule
DVS-G & S-LISTA
& $-12.50 \pm 0.04$
& $\mathbf{94.72 \pm 0.75}$
& $\mathbf{65{,}536}$ \\

DVS-G & Vanilla SNN
& $-10.08 \pm 0.21$
& $90.90 \pm 1.05$
& $\mathbf{65{,}536}$ \\

DVS-G & ANN LISTA
& $-14.70 \pm 0.14$
& $94.10 \pm 0.78$
& $524{,}288$ \\

DVS-G & ANN LSTM
& $\mathbf{-15.04 \pm 0.04}$
& $94.51 \pm 0.62$
& $524{,}288$ \\
\midrule
SHD & S-LISTA
& $\mathbf{-16.66 \pm 0.10}$
& $89.76 \pm 0.33$
& $\mathbf{6{,}400}$ \\

SHD & Vanilla SNN
& $-12.17 \pm 0.29$
& $89.19 \pm 1.02$
& $\mathbf{6{,}400}$ \\

SHD & ANN LISTA
& $-16.45 \pm 0.04$
& $92.55 \pm 1.09$
& $51{,}200$ \\

SHD & ANN LSTM
& $-16.19 \pm 0.10$
& $\mathbf{93.03 \pm 0.62}$
& $51{,}200$ \\
\bottomrule
\end{tabular*}
\endgroup
\end{table*}

\begin{figure*}[htbp]
\centering
\includegraphics[width=\textwidth]{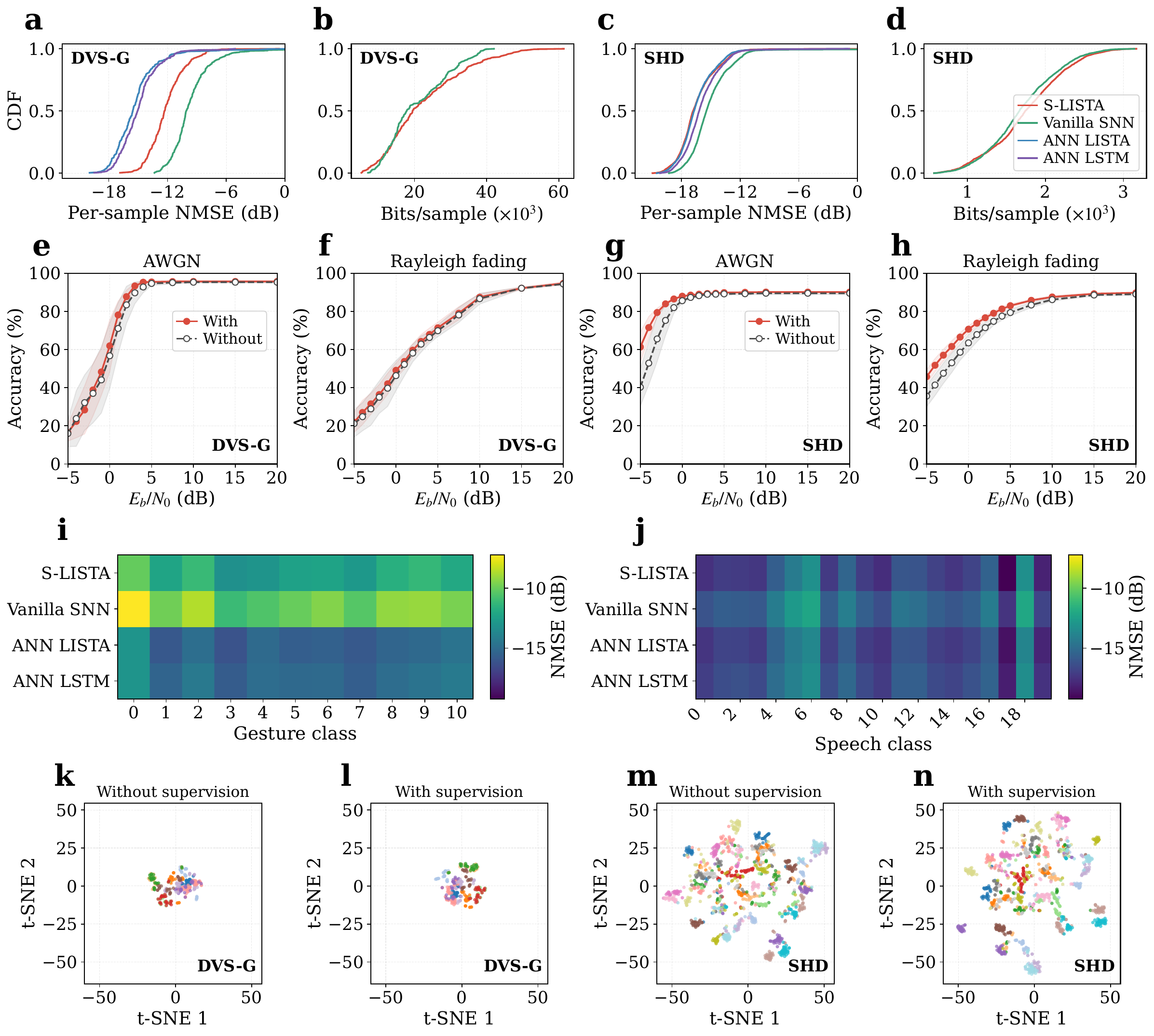}
\caption{\textbf{Sample-wise performance and the effect of task supervision.}
\textbf{a,b}, Empirical cumulative distribution functions (CDFs) of per-sample NMSE and AER bits per sample on DVS-G.
\textbf{c,d}, Corresponding distributions on SHD.
\textbf{e,f}, Downstream task accuracy with and without task supervision on DVS-G over AWGN and Rayleigh fading, respectively.
\textbf{g,h}, Corresponding results on SHD.
\textbf{i,j}, Class-wise reconstruction NMSE on DVS-G and SHD, respectively.
\textbf{k,l}, t-distributed stochastic neighbor embedding (t-SNE) visualizations of S-LISTA sparse codes on DVS-G without and with task supervision, respectively.
\textbf{m,n}, Corresponding visualizations on SHD.
Colours distinguish classes.
NMSE distributions, class-wise NMSE, and t-SNE visualizations use full-vector transmission under Rayleigh fading at 20~dB.
Accuracy scans also use full-vector transmission; bit-count distributions use AER.
All models are trained over AWGN at 10~dB. In panels \textbf{e--h}, curves and shaded bands indicate means and one standard deviation across five training seeds, respectively.}
\label{fig:event_distributions}
\end{figure*}

\begin{table*}[t]
\centering
\caption{\textbf{Core ablation results under Rayleigh fading at 20~dB.}
Results use block-based AER transmission and are reported as means $\pm$ standard deviations over five independent seeds.}
\label{tab:event_ablation}
\begingroup
\footnotesize
\setlength{\tabcolsep}{5pt}
\renewcommand{\arraystretch}{1.06}
\begin{tabular*}{\textwidth}{@{\extracolsep{\fill}}llccc@{}}
\toprule
Dataset & Setting & NMSE (dB) & Accuracy (\%) & Bits/sample \\
\midrule
DVS-G & Without temperature relaxation
& $-8.88 \pm 0.26$
& $87.64 \pm 1.27$
& $59{,}262 \pm 17{,}870$ \\

DVS-G & Without rate regularization
& $-11.70 \pm 0.13$
& $91.88 \pm 1.00$
& $112{,}403 \pm 49{,}419$ \\

DVS-G & Without task supervision
& $-11.54 \pm 0.09$
& $92.50 \pm 0.76$
& $20{,}651 \pm 1{,}043$ \\

DVS-G & S-LISTA with AER
& $-11.51 \pm 0.07$
& $92.85 \pm 0.63$
& $20{,}252 \pm 1{,}538$ \\
\midrule
SHD & Without temperature relaxation
& $-15.31 \pm 0.16$
& $89.29 \pm 0.44$
& $2{,}748 \pm 78$ \\

SHD & Without rate regularization
& $-15.81 \pm 0.08$
& $89.55 \pm 0.78$
& $3{,}919 \pm 83$ \\

SHD & Without task supervision
& $-15.28 \pm 0.07$
& $87.39 \pm 1.03$
& $1{,}418 \pm 28$ \\

SHD & S-LISTA with AER
& $-15.05 \pm 0.09$
& $88.58 \pm 0.43$
& $1{,}769 \pm 19$ \\
\bottomrule
\end{tabular*}
\endgroup
\end{table*}

\subsection{Sample-wise performance and ablation analysis}

We further evaluate performance across individual samples and examine the contributions of the framework's main components through ablation experiments. The NMSE distributions in Figs.~\ref{fig:event_distributions}a,c show that S-LISTA's reconstruction advantage over the vanilla SNN extends across a broad range of samples. On DVS-G, its reconstruction errors generally lie between those of the ANN methods and the vanilla SNN, whereas on SHD its distribution is close to that of ANN LISTA. The class-wise results in Figs.~\ref{fig:event_distributions}i,j show a similar trend. The AER bit-count distributions in Figs.~\ref{fig:event_distributions}b,d further show that the transmitted payload varies across samples with spike activity. The two SNN methods have similar payload distributions on DVS-G, while S-LISTA tends to require fewer bits on SHD.

Table~\ref{tab:event_ablation} assesses the contributions of temperature relaxation, rate regularization, and task supervision. Without temperature relaxation, NMSE increases by 2.63~dB and downstream task accuracy decreases by 5.21 percentage points on DVS-G, whereas the performance changes on SHD are smaller. Without rate regularization, transmitted bits increase by approximately $455\%$ on DVS-G and $122\%$ on SHD, with NMSE improvements of only 0.19 and 0.76~dB, respectively. Task supervision improves downstream task performance: its removal reduces mean accuracy by 0.35 and 1.19 percentage points on DVS-G and SHD, respectively.

Task supervision also improves robustness under challenging channel conditions, as shown in Figs.~\ref{fig:event_distributions}e--h. On DVS-G, it increases downstream task accuracy from $83.40\%$ to $87.71\%$ at 2~dB over AWGN, a gain of 4.31 percentage points. On SHD, the supervised model achieves higher mean accuracy at every tested SNR under both AWGN and Rayleigh fading. At 2~dB under Rayleigh fading, accuracy increases from $71.48\%$ to $76.66\%$, a gain of 5.19 percentage points.

\begin{table*}[t]
\centering
\caption{\textbf{Computational cost and estimated energy consumption per sample.}}
\label{tab:event_computational_cost}
\begingroup
\footnotesize
\setlength{\tabcolsep}{3pt}
\renewcommand{\arraystretch}{1.08}
\begin{tabular*}{\textwidth}{@{\extracolsep{\fill}}llrrrrrr@{}}
\toprule
& & \multicolumn{3}{c}{Sensing}
& \multicolumn{3}{c}{Reconstruction} \\
\cmidrule(lr){3-5}\cmidrule(lr){6-8}
Dataset & Method
& MAC (MOP) & AC (MOP) & Energy ($\mu$J)
& MAC (MOP) & AC (MOP) & Energy ($\mu$J) \\
\midrule
DVS-G & S-LISTA
& 0 & 88.9 & 80.1
& 0 & 4{,}350 & 3{,}910 \\

& Vanilla SNN
& 0 & 126 & 113
& 1{,}210 & 1{,}690 & 7{,}070 \\

& ANN LISTA
& 340 & 0 & 1{,}560
& 19{,}400 & 0 & 89{,}300 \\

& ANN LSTM
& 340 & 0 & 1{,}560
& 9{,}970 & 0 & 45{,}800 \\
\midrule
SHD & S-LISTA
& 0 & 2.11 & 1.90
& 0 & 5.60 & 5.04 \\

& Vanilla SNN
& 0 & 2.11 & 1.90
& 0 & 23.7 & 21.4 \\

& ANN LISTA
& 4.48 & 0 & 20.6
& 90.2 & 0 & 415 \\

& ANN LSTM
& 4.48 & 0 & 20.6
& 39.3 & 0 & 181 \\
\bottomrule
\end{tabular*}

\smallskip
\begin{minipage}{\textwidth}
\footnotesize
Operation counts cover all time steps of each input sample.
One MOP denotes $10^6$ operations, with each multiply--accumulate
(MAC) or accumulate (AC) counted as one operation.
Energy is estimated using 4.6~pJ per MAC and 0.9~pJ per AC.
The final dictionary readout, communication, memory access,
and downstream task evaluation are excluded.
\end{minipage}
\endgroup
\end{table*}

\subsection{Computational cost and estimated energy}

We assess the computational cost and estimated energy consumption of the framework in Table~\ref{tab:event_computational_cost}. At the sensing stage, the S-LISTA-based neuromorphic compressive sensing framework reduces estimated energy consumption by factors of 19.5 on DVS-G and 10.9 on SHD relative to the ANN methods. At the reconstruction stage, S-LISTA reduces estimated energy consumption relative to the vanilla SNN by factors of 1.81 and 4.24, respectively. These savings support the connection between signal sparsity and computational sparsity established through the S-LISTA structure. The combined sensing and reconstruction energy is approximately $3.99$~mJ per sample on DVS-G and $6.93\,\mu\mathrm{J}$ on SHD, corresponding to reductions of approximately $95.6\%$ and $98.4\%$ relative to ANN LISTA. Together with the previous results, these estimates show that the framework reduces energy consumption while maintaining competitive reconstruction and downstream task performance, making it well suited to resource-constrained edge devices.

\section{Discussion}
\label{sec:discussion}

This work presents a neuromorphic compressive sensing framework that extends conventional dense CS to an event-driven paradigm. The framework represents compressed measurements and reconstruction updates as spikes, allowing communication and computational costs to depend on event activity. The proposed S-LISTA preserves the LISTA update structure and constructs a sparse code by accumulating spike updates across layers. The framework uses joint training of the measurement network, reconstruction network, and dictionary to improve end-to-end reconstruction performance, with task supervision to retain information needed for downstream tasks.

Experiments on synthetic sparse signals, ECG, and neuromorphic event datasets highlight three main benefits of the proposed framework. First, S-LISTA enables energy-efficient reconstruction, achieving competitive reconstruction quality with substantially lower estimated energy than ANN baselines. Second, joint learning of the sensing matrix and dictionary improves reconstruction, with the dictionary adapting to the sparse codes produced by spike-based recovery. Third, on visual and auditory event data, the complete framework reduces transmission cost with modest trade-offs in reconstruction and downstream task performance. Evaluations over AWGN and Rayleigh fading further demonstrate robustness under challenging channel conditions, including fading not encountered during training.

These properties are particularly relevant to applications that require both signal reconstruction and downstream analysis, such as accelerated magnetic resonance imaging~\cite{10578304}, biological microscopy~\cite{dai2025implicit,lee2025adaptive}, and event-based vision~\cite{rebecq2019events}. For edge deployments in these scenarios, limited bandwidth and energy make efficient transmission and reconstruction particularly important. By linking sparse signal representations with sparse SNN activity, neuromorphic compressive sensing provides a promising paradigm for reconstruction on resource-constrained edge devices.

\section{Methods}
\label{sec:methods}

\subsection{Neuromorphic sensing based on SNNs}
\label{subsec:spike_sensing}

The transmitter maps a source sequence $\{\mathbf{x}_t\}_{t=1}^{T}$, with $\mathbf{x}_t\in\mathbb{R}^{N}$, to low-dimensional measurement spikes. A learnable mapping $\boldsymbol{\Phi}:\mathbb{R}^{N}\rightarrow\mathbb{R}^{M}$ generates the measurement current, with $M<N$. LIF neurons convert this current into binary spikes~\cite{gerstner2002spiking}:
\begin{equation}
\begin{aligned}
\bar{\mathbf{u}}^{\rm tx}_t
&= \rho_{\rm tx}\mathbf{u}^{\rm tx}_{t-1}
+ \boldsymbol{\Phi}(\mathbf{x}_t), \\
\mathbf{s}^{\rm tx}_t
&= H\!\left(\bar{\mathbf{u}}^{\rm tx}_t-\vartheta_{\rm tx}\right), \\
\mathbf{u}^{\rm tx}_t
&= \bar{\mathbf{u}}^{\rm tx}_t
-\vartheta_{\rm tx}\mathbf{s}^{\rm tx}_t,
\end{aligned}
\label{eq:spike_sensing}
\end{equation}
where $\bar{\mathbf{u}}^{\rm tx}_t$ and $\mathbf{u}^{\rm tx}_t$ are the membrane potentials before and after reset, respectively, with $\mathbf{u}^{\rm tx}_0=\mathbf{0}$. The parameters $\rho_{\rm tx}\in[0,1)$ and $\vartheta_{\rm tx}>0$ denote the decay factor and firing threshold. The Heaviside function $H$ acts element-wise, yielding $\mathbf{s}^{\rm tx}_t\in\{0,1\}^{M}$.

The final line of Eq.~\eqref{eq:spike_sensing} implements soft reset~\cite{Han_2020_CVPR}. The residual membrane potential carries measurement information to subsequent time steps. Measurement spikes are encoded for transmission to the receiver.

\subsection{Wireless transmission and block-based AER}
\label{subsec:aer_transmission}

Continuous measurements are uniformly quantized to $q$ bits within a prescribed range and converted into a bit sequence. Measurement spikes are represented either as full binary vectors or using AER~\cite{lichtsteiner2008128}. Full binary vectors include the spike state at every measurement coordinate, whereas AER records only the addresses of active coordinates. The resulting bits are transmitted using binary phase-shift keying (BPSK) over an AWGN or Rayleigh fading channel. The received signal and coherent hard decision are
\begin{equation}
\begin{aligned}
r_j
&= h\sqrt{E_b}(2b_j-1)+n_j, \\
\hat{b}_j
&= \mathbf{1}\!\left\{\operatorname{Re}(h^*r_j)\geq 0\right\},
\end{aligned}
\label{eq:wireless_channel}
\end{equation}
where $b_j,\hat{b}_j\in\{0,1\}$ are the transmitted and received bits, $j$ indexes the transmitted bits within a sample, and $E_b$ is the energy per bit. The indicator function $\mathbf{1}\{\cdot\}$ equals one when its condition holds and zero otherwise. The noise samples are independent, with $n_j\sim\mathcal{CN}(0,N_0)$, where $N_0$ is the variance of the complex noise. The channel coefficient is $h=1$ for AWGN and $h\sim\mathcal{CN}(0,1)$ for Rayleigh fading. Under quasi-static fading, $h$ remains constant across all transmissions of a sample and is independent between samples. The receiver has perfect channel state information, whereas the transmitter has none.

We simulate the hard decisions using the conditional BPSK error probability:
\begin{equation}
\begin{aligned}
p_{\rm e}(h)
&= \frac{1}{2}\operatorname{erfc}
\!\left(\sqrt{\gamma|h|^2}\right), \\
\hat{b}_j
&= b_j\oplus e_j,
\qquad
e_j\mid h\sim\operatorname{Bernoulli}\!\left(p_{\rm e}(h)\right),
\end{aligned}
\label{eq:conditional_bit_error}
\end{equation}
where $\gamma=E_b/N_0$, $\operatorname{erfc}$ is the complementary error function, and $\oplus$ denotes exclusive OR. Bit errors are sampled independently conditional on $h$. This procedure gives the same conditional bit-error distribution as Eq.~\eqref{eq:wireless_channel}. The receiver converts the detected bits back into quantized measurements or decodes the spike states and AER addresses to recover the binary measurement vector.

To limit coordinate displacements caused by local-address errors, we divide the $M$ measurement coordinates into blocks of size $Q$, with $Q$ dividing $M$. Each non-empty block transmits its block address once, followed by the local addresses of its active events. The AER payload per sample is
\begin{equation}
B^{\rm AER}
=
\sum_{t=1}^{T}
\left[
N_t^{\rm blk}
\left\lceil\log_2\frac{M}{Q}\right\rceil
+
\left\|\mathbf{s}_t^{\rm tx}\right\|_0
\left\lceil\log_2 Q\right\rceil
\right],
\label{eq:aer_payload}
\end{equation}
where $N_t^{\rm blk}$ is the number of non-empty blocks at time step $t$, and $\|\mathbf{s}_t^{\rm tx}\|_0$ counts its active events. The two terms count block-address and local-address bits, respectively. The payload excludes framing and synchronization overhead. AER reduces the payload when these address bits require fewer bits than full-vector transmission. A local-address error changes an event's position within the decoded block, whereas a block-address error redirects all events associated with that block. Events decoded to the same coordinate are merged into one spike.

\subsection{S-LISTA architecture}
\label{subsec:spiking_lista}

At the receiver, S-LISTA reconstructs the signals from the received measurement vector, $\widetilde{\mathbf{s}}_t^{\rm rx}$. It follows the measurement embedding and residual correction structure of unfolded sparse recovery~\cite{sprechmann2015learning,chen2018theoretical}, with spike updates accumulated over $L$ layers. Each layer's membrane potential is initialized as $\mathbf{u}_0^{(\ell)}=\mathbf{0}$ at the start of a sequence. At each time step, the sparse code is initialized as $\mathbf{z}_t^{(0)}=\mathbf{0}$, and the received measurements are embedded as
\begin{equation}
\mathbf{I}_t^{(1)}
=
\mathbf{P}_{\rm s}\widetilde{\mathbf{s}}_t^{\rm rx},
\label{eq:slista_embedding}
\end{equation}
where $\mathbf{P}_{\rm s}$ is the learnable measurement-embedding operator and $\mathbf{I}_t^{(1)}$ is the input current to the first layer.

Each layer integrates its input current, generates signed spikes, and updates the sparse code:
\begin{equation}
\begin{aligned}
\bar{\mathbf{u}}_t^{(\ell)}
&=
\rho_{\rm rx}\mathbf{u}_{t-1}^{(\ell)}
+\mathbf{I}_t^{(\ell)}, \\
\boldsymbol{\xi}_t^{(\ell)}
&=
H\!\left(\bar{\mathbf{u}}_t^{(\ell)}
-\vartheta_{\rm rx}^{(\ell)}\right)
-
H\!\left(-\bar{\mathbf{u}}_t^{(\ell)}
-\vartheta_{\rm rx}^{(\ell)}\right), \\
\mathbf{u}_t^{(\ell)}
&=
\bar{\mathbf{u}}_t^{(\ell)}
-\vartheta_{\rm rx}^{(\ell)}\boldsymbol{\xi}_t^{(\ell)}, \\
\mathbf{z}_t^{(\ell)}
&=
\mathbf{z}_t^{(\ell-1)}
+\boldsymbol{\xi}_t^{(\ell)}, \\
\mathbf{I}_t^{(\ell+1)}
&=
\mathbf{I}_t^{(1)}
-\mathbf{G}_{\rm s}^{(\ell)}\mathbf{z}_t^{(\ell)},
\qquad \ell<L,
\end{aligned}
\label{eq:spiking_lista}
\end{equation}
where $\ell=1,\ldots,L$ indexes the unfolded layers and $t$ indexes time steps. The variables $\bar{\mathbf{u}}_t^{(\ell)}$ and $\mathbf{u}_t^{(\ell)}$ are the membrane potentials before and after soft reset, respectively. The parameters $\rho_{\rm rx}\in[0,1)$ and $\vartheta_{\rm rx}^{(\ell)}>0$ denote the decay factor and firing threshold. Each entry of $\boldsymbol{\xi}_t^{(\ell)}$ belongs to $\{-1,0,1\}$. The learnable operator $\mathbf{G}_{\rm s}^{(\ell)}$ maps the accumulated sparse code to the residual current for the next layer.

After the final layer, the residual membrane potential provides a continuous correction at the nonzero coordinates of the accumulated code. An output soft threshold is then applied before dictionary reconstruction:
\begin{equation}
\begin{aligned}
\mathbf{z}_t
&=
\mathcal{S}_{\lambda_{\rm out}}\!\left(
\mathbf{z}_t^{(L)}
+
\mathbf{1}\!\left[\mathbf{z}_t^{(L)}\neq0\right]
\odot
\frac{\mathbf{u}_t^{(L)}}{\vartheta_{\rm rx}^{(L)}}
\right), \\
\widehat{\mathbf{x}}_t
&=
\boldsymbol{\Psi}\mathbf{z}_t,
\end{aligned}
\label{eq:slista_readout}
\end{equation}
where $\boldsymbol{\Psi}$ is the reconstruction dictionary, $\mathbf{1}[\cdot]$ acts element-wise, and $\odot$ denotes element-wise multiplication. The soft-thresholding operator acts element-wise as $\mathcal{S}_{\lambda}(a)=\operatorname{sign}(a)\max(|a|-\lambda,0)$, with output threshold $\lambda_{\rm out}\geq0$. Setting $\lambda_{\rm out}=0$ leaves the corrected code unchanged. The pseudocode of S-LISTA is provided in Supplementary Note~1. Supplementary Note~3, Section 3.3 establishes a recovery error bound given the true support, an amplitude bound, and a noise bound. Supplementary Note~3, Section~3.4 formalizes the connection between signal sparsity and computational sparsity through an upper bound on the S-LISTA firing rate.

\subsection{Training objective and gradient estimation}
\label{subsec:joint_learning}

The framework uses joint training to optimize the sensing network, S-LISTA, the dictionary, and an auxiliary classifier. The classifier takes the final sparse-code sequence $\mathbf{Z}=\{\mathbf{z}_t\}_{t=1}^{T}$ as input and provides task supervision. The training objective combines reconstruction error, sparse-code regularization, measurement spike rate, and task supervision:
\begin{equation}
\begin{aligned}
\mathcal{L}_{\rm joint}
=
\mathbb{E}_{\mathbf{X},y,\mathbf{n}}
\Bigg[
&\frac{1}{T}\sum_{t=1}^{T}
\Bigg(
\frac{1}{2}
\left\|\widehat{\mathbf{x}}_t-\mathbf{x}_t\right\|_2^2
+
\frac{\lambda_{\rm s}}{L}
\sum_{\ell=1}^{L}
\left\|\mathbf{z}_t^{(\ell)}\right\|_1 \\
&\qquad\qquad
+
\frac{\lambda_{\rm r}}{M}
\sum_{m=1}^{M}
\pi_{t,m}
\Bigg)
+
\frac{\lambda_{\rm sem}}{\log C}
\operatorname{CE}\!\left(
g_{\boldsymbol{\omega}}(\mathbf{Z}),y
\right)
\Bigg],
\label{eq:joint_training_objective}
\end{aligned}
\end{equation}
where $\mathbf{X}=\{\mathbf{x}_t\}_{t=1}^{T}$ is the input sequence, and the expectation is over training samples $(\mathbf{X},y)$ and AWGN $\mathbf{n}$. $\lambda_{\rm s}$, $\lambda_{\rm r}$, and $\lambda_{\rm sem}$ are the weight coefficients of the sparse-code, spike-rate, and task terms, respectively. The quantity $\pi_{t,m}$ is the relaxed firing probability for measurement coordinate $m$ at time step $t$. The function $g_{\boldsymbol{\omega}}$ is the auxiliary classifier with parameters $\boldsymbol{\omega}$, $\operatorname{CE}$ denotes cross-entropy, $y$ is the class label, and $C$ is the number of classes. The task term acts on the sparse codes without passing through the dictionary. Reported downstream task performance is evaluated separately using reconstructed signals. The rate term constrains measurement activity, while the AER payload is computed using Eq.~\eqref{eq:aer_payload}.

Hard spike generation is non-differentiable. Therefore, the sensing network uses a continuous approximation to estimate gradients during training:
\begin{equation}
\begin{aligned}
\pi_{t,m}
&=
\sigma\!\left(
\frac{\bar{u}_{t,m}^{\rm tx}-\vartheta_{\rm tx}}
{\tau_{e_{\rm tr}}}
\right), \\
\tau_{e_{\rm tr}}
&=
\tau_{\rm init}
\left(
\frac{\tau_{\rm final}}{\tau_{\rm init}}
\right)^{e_{\rm tr}/(E-1)},
\end{aligned}
\label{eq:signed_spike_relaxation}
\end{equation}
where $\sigma$ is the sigmoid function. The index $e_{\rm tr}=0,\ldots,E-1$ denotes the training epoch, and $E>1$ is the total number of epochs. The temperature decreases from $\tau_{\rm init}$ to $\tau_{\rm final}$, making the approximation closer to the hard threshold function. The transmitter uses hard spikes in the forward pass and gradients of the relaxed probabilities in the backward pass under the straight-through estimator~\cite{bengio2013estimating,maddison2016concrete,jang2016categorical}. S-LISTA also uses surrogate gradients for its hard spike functions.

For full-vector binary transmission, the backward pass uses the probability that the received bit equals $1$:
\begin{equation}
p_j^{\rm rx}
=
(1-p_{\rm e})\bar{b}_j
+
p_{\rm e}(1-\bar{b}_j)
=
p_{\rm e}+(1-2p_{\rm e})\bar{b}_j,
\label{eq:awgn_decision_relaxation}
\end{equation}
where $\bar{b}_j\in[0,1]$ is the relaxed probability that the transmitted bit equals $1$, and $p_j^{\rm rx}$ is the probability that the received bit equals $1$. The bit-error probability $p_{\rm e}$ is given by Eq.~\eqref{eq:conditional_bit_error} with $h=1$ for AWGN training. For block-based AER, the backward pass uses a differentiable address mapping based on the error probabilities of block and local addresses. The forward pass and inference retain hard address decoding and hard bit decisions.

\subsection{Experimental procedures and analytical energy estimates}
\label{subsec:experimental_energy}

We evaluate S-LISTA on synthetic sparse signals, ECG signals, and neuromorphic event streams. All experiments were conducted on NVIDIA H100 NVL GPUs. Comparisons with ANN and vanilla SNN baselines use matched measurement dimensions and channel conditions. Dataset preparation, network architectures, and training hyperparameters are provided in Supplementary Note~2.

All methods are implemented in PyTorch~\cite{paszke2019pytorch} and trained using Adam with cosine learning-rate annealing. The synthetic experiments use online-generated training data and separate validation and test sets. For ECG, DS1 is divided into training and validation subsets, with approximately 20\% reserved for validation and model selection, while DS2 is used exclusively for testing. Following previous practice~\cite{Fang_2021_ICCV,baronig2025advancing}, we use the provided training and test splits for DVS-G and SHD, with the test splits used for both model selection and evaluation. Since this practice is not methodologically rigorous~\cite{baronig2025advancing}, we additionally report results with independent validation sets in Supplementary Note~5. Wireless transmission experiments use AWGN during training and include both AWGN and Rayleigh fading during evaluation. All experiments are repeated with five independent random seeds.

We estimate energy consumption from operation counts across all time steps of each sample. Each MAC operation or accumulate (AC) operation is counted once. Dense matrix computations contribute MACs, whereas spike-triggered additions of synaptic weights contribute ACs. The estimated energy per sample is
\begin{equation}
E_{\rm op}
=
\epsilon_{\rm MAC}N_{\rm MAC}
+
\epsilon_{\rm AC}N_{\rm AC},
\label{eq:operation_energy}
\end{equation}
where $N_{\rm MAC}$ and $N_{\rm AC}$ are the operation counts per sample. We use $\epsilon_{\rm MAC}=4.6$~pJ and $\epsilon_{\rm AC}=0.9$~pJ, corresponding to representative 32-bit floating-point operation costs~\cite{horowitz20141}. The estimates cover sensing and reconstruction but exclude the final dictionary readout, communication, memory access, downstream task evaluation, and hardware control costs.

\section{Data availability}
\label{sec:Data availability}

The MIT-BIH Arrhythmia Database is publicly available at \url{https://physionet.org/content/mitdb/1.0.0/} under the ODC BY 1.0 license. The DVS-Gesture dataset is publicly available at \url{https://ibm.biz/EventCameraData} under the CC BY 4.0 license. The Spiking Heidelberg Digits dataset is publicly available at \url{https://zenkelab.org/resources/spiking-heidelberg-datasets-shd/} under the CC BY 4.0 license. Code for generating the synthetic dataset is available at \url{https://github.com/zerufang01-max/neuromorphic-compressive-sensing}.

\section{Code availability}
\label{sec:Code availability}

The source code is publicly available at \url{https://github.com/zerufang01-max/neuromorphic-compressive-sensing}.

\section*{Author contributions}
Z.F. conceived the idea. Z.F. and Y.L. developed the methodology.
Z.F. conducted the experiments and wrote the manuscript.
Y.L. guided the experiments and revised the manuscript.
G.Y.L. supervised the research and revised the manuscript.
All authors read and approved the final manuscript.

\section*{Competing interests}
The authors declare no competing interests.

\end{document}


\begin{center}
{\Large \textbf{Supplementary Information for}}\\[0.5em]
{\LARGE \textbf{Event-driven signal reconstruction through neuromorphic compressive sensing}}\\[1em]
{\large Zeru Fang et al.}
\end{center}

\section*{Supplementary Note 1: S-LISTA algorithm}

The spike-driven learned iterative shrinkage-thresholding algorithm (S-LISTA) retains three operations from the learned iterative shrinkage-thresholding algorithm (LISTA)~\cite{gregor2010learning}: measurement
embedding, residual correction, and sparse-code updating. S-LISTA takes the received measurement sequence
$\{\widetilde{\mathbf{s}}^{\rm rx}_t\}_{t=1}^{T}$ as input and uses $L$
unfolded layers. The membrane states are initialized as
$\mathbf{u}_0^{(\ell)}=\mathbf{0}$ at the start of each sequence and retained
across time steps. At each time step, the accumulated sparse code starts
from $\mathbf{z}_t^{(0)}=\mathbf{0}$.

\textbf{Measurement embedding.}
The received measurements are mapped to the sparse-code dimension as
\begin{equation}
    \mathbf{I}_{t}^{(1)}
    =\mathbf{P}_{\rm s}\widetilde{\mathbf{s}}^{\rm rx}_{t},
    \label{eq:supp_embedding}
\end{equation}
where $\mathbf{P}_{\rm s}$ is the learnable measurement-embedding operator
and $\mathbf{I}_{t}^{(1)}$ is the input current to the first layer.
This corresponds to the measurement mapping in LISTA. The embedding is
computed once per time step and reused by subsequent layers.

\textbf{Sparse-code update.}
For $\ell=1,\ldots,L$, the layer integrates its input current and generates
a signed spike increment:
\begin{align}
    \bar{\mathbf{u}}_{t}^{(\ell)}
    &=\rho_{\rm rx}\mathbf{u}_{t-1}^{(\ell)}
      +\mathbf{I}_{t}^{(\ell)},
    \label{eq:supp_membrane}\\
    \boldsymbol{\xi}_{t}^{(\ell)}
    &=H\!\left(\bar{\mathbf{u}}_{t}^{(\ell)}
      -\vartheta_{\rm rx}^{(\ell)}\right)
      -H\!\left(-\bar{\mathbf{u}}_{t}^{(\ell)}
      -\vartheta_{\rm rx}^{(\ell)}\right),
    \label{eq:supp_signed_spike}
\end{align}
where $\rho_{\rm rx}\in[0,1)$ is the membrane decay factor,
$\vartheta_{\rm rx}^{(\ell)}>0$ is the firing threshold, and $H$ is the
Heaviside function applied element-wise. Each entry of
$\boldsymbol{\xi}_{t}^{(\ell)}$ belongs to $\{-1,0,1\}$.
Positive and negative threshold crossings increase and decrease the
corresponding code coefficient, respectively. The signed soft reset and
code accumulation are
\begin{align}
    \mathbf{u}_{t}^{(\ell)}
    &=\bar{\mathbf{u}}_{t}^{(\ell)}
      -\vartheta_{\rm rx}^{(\ell)}\boldsymbol{\xi}_{t}^{(\ell)},
    \label{eq:supp_soft_reset}\\
    \mathbf{z}_{t}^{(\ell)}
    &=\mathbf{z}_{t}^{(\ell-1)}+\boldsymbol{\xi}_{t}^{(\ell)},
    \label{eq:supp_code_accumulation}
\end{align}
where $\bar{\mathbf{u}}_{t}^{(\ell)}$ and $\mathbf{u}_{t}^{(\ell)}$ are
the membrane potentials before and after reset, and
$\mathbf{z}_{t}^{(\ell)}$ is the code accumulated through layer $\ell$.
Spike accumulation replaces the continuous soft-threshold update in ANN LISTA. The post-reset membrane potential retains
the part of the input that has not been expressed by the spike increment.

\textbf{Residual correction.}
For $\ell<L$, the next layer receives the measurement embedding corrected
by the current sparse code:
\begin{equation}
    \mathbf{I}_{t}^{(\ell+1)}
    =\mathbf{I}_{t}^{(1)}
      -\mathbf{G}_{\rm s}^{(\ell)}\mathbf{z}_{t}^{(\ell)},
    \label{eq:supp_residual_correction}
\end{equation}
where $\mathbf{G}_{\rm s}^{(\ell)}$ is a learnable residual-correction
operator. This corresponds to the code-dependent correction in LISTA.
Each layer therefore uses the code accumulated by the previous layers to
form its input current. Code accumulation proceeds across layers within
each time step, while each layer's membrane state persists across time.

\textbf{Continuous correction and reconstruction.}
After the final layer, the accumulated code selects the coordinates for
continuous correction:
\begin{equation}
    \mathbf{m}_{t}
    =\mathbf{1}\!\left[\mathbf{z}_{t}^{(L)}\neq0\right],
    \label{eq:supp_support_mask}
\end{equation}
where $\mathbf{1}[\cdot]$ equals one when its condition holds and zero
otherwise, and acts element-wise. The final post-reset membrane potential
refines the coefficient amplitudes on this support. The corrected code is
soft-thresholded and mapped through the dictionary:
\begin{align}
    \mathbf{z}_{t}
    &=\mathcal{S}_{\lambda_{\rm out}}\!\left(
      \mathbf{z}_{t}^{(L)}+
      \mathbf{m}_{t}\odot
      \frac{\mathbf{u}_{t}^{(L)}}{\vartheta_{\rm rx}^{(L)}}
      \right),
    \label{eq:supp_readout}\\
    \widehat{\mathbf{x}}_{t}
    &=\boldsymbol{\Psi}\mathbf{z}_{t},
    \label{eq:supp_reconstruction}
\end{align}
where $\boldsymbol{\Psi}$ is the dictionary, $\odot$ denotes
element-wise multiplication, and
$\mathcal{S}_{\lambda}(a)=\operatorname{sign}(a)\max(|a|-\lambda,0)$
acts element-wise. The output threshold satisfies $\lambda_{\rm out}\geq0$.

\begin{algorithm}[htbp]
\caption{S-LISTA inference}
\label{alg:spiking_lista_forward}
\begin{algorithmic}[1]
\STATE \textbf{Input:} Received measurements
$\{\widetilde{\mathbf{s}}_{t}^{\rm rx}\}_{t=1}^{T}$;
number of layers $L$; measurement-embedding operator $\mathbf{P}_{\rm s}$;
residual-correction operators $\{\mathbf{G}_{\rm s}^{(\ell)}\}_{\ell=1}^{L-1}$;
dictionary $\boldsymbol{\Psi}$; decay factor $\rho_{\rm rx}$;
firing thresholds $\{\vartheta_{\rm rx}^{(\ell)}\}_{\ell=1}^{L}$;
output threshold $\lambda_{\rm out}$.
\STATE Initialize $\mathbf{u}_{0}^{(\ell)}=\mathbf{0}$ for $\ell=1,\ldots,L$.
\FOR{$t=1,\ldots,T$}
    \STATE $\mathbf{I}_{t}^{(1)}\gets
    \mathbf{P}_{\rm s}\widetilde{\mathbf{s}}_{t}^{\rm rx}$
    \STATE $\mathbf{z}_{t}^{(0)}\gets\mathbf{0}$
    \FOR{$\ell=1,\ldots,L$}
        \STATE $\bar{\mathbf{u}}_{t}^{(\ell)}\gets
        \rho_{\rm rx}\mathbf{u}_{t-1}^{(\ell)}+\mathbf{I}_{t}^{(\ell)}$
        \STATE Generate $\boldsymbol{\xi}_{t}^{(\ell)}$ using
        Eq.~\eqref{eq:supp_signed_spike}.
        \STATE $\mathbf{u}_{t}^{(\ell)}\gets
        \bar{\mathbf{u}}_{t}^{(\ell)}
        -\vartheta_{\rm rx}^{(\ell)}\boldsymbol{\xi}_{t}^{(\ell)}$
        \STATE $\mathbf{z}_{t}^{(\ell)}\gets
        \mathbf{z}_{t}^{(\ell-1)}+\boldsymbol{\xi}_{t}^{(\ell)}$
        \IF{$\ell<L$}
            \STATE $\mathbf{I}_{t}^{(\ell+1)}\gets
            \mathbf{I}_{t}^{(1)}
            -\mathbf{G}_{\rm s}^{(\ell)}\mathbf{z}_{t}^{(\ell)}$
        \ENDIF
    \ENDFOR
    \STATE $\mathbf{m}_{t}\gets
    \mathbf{1}[\mathbf{z}_{t}^{(L)}\neq0]$
    \STATE $\mathbf{z}_{t}\gets
    \mathcal{S}_{\lambda_{\rm out}}\!\left(
    \mathbf{z}_{t}^{(L)}+\mathbf{m}_{t}\odot
    \mathbf{u}_{t}^{(L)}/\vartheta_{\rm rx}^{(L)}\right)$
    \STATE $\widehat{\mathbf{x}}_{t}\gets\boldsymbol{\Psi}\mathbf{z}_{t}$
\ENDFOR
\STATE \textbf{Output:} Reconstructed sequence
$\{\widehat{\mathbf{x}}_{t}\}_{t=1}^{T}$ and sparse codes
$\{\mathbf{z}_{t}\}_{t=1}^{T}$.
\end{algorithmic}
\end{algorithm}

\section*{Supplementary Note 2: Experimental settings and evaluation}
\label{supp:experimental_protocols}

\subsection*{Datasets and network configurations}

\textbf{Synthetic signals.}
For the energy--reconstruction comparisons, signals have 256 coordinates, with $s\in\{14,28,42\}$ nonzero entries at
random positions, magnitudes $\mathcal{U}[0,4)$, and equally probable signs.
A fixed Gaussian matrix with unit-norm columns produces 141 measurements.
Training signals are generated online, with no fixed training-set size.
Independent sets of 1,536, 6,000, and
10,000 signals are used for checkpoint selection, hyperparameter selection, and
testing, respectively. ANN depths are 2--5; S-LISTA depths are 5, 10, 15, and 20.
Depth is varied to compare reconstruction error against estimated energy.
The layerwise example in Fig.~2e uses $s=14$ and five unfolded layers, with the lowest-NMSE example selected across 6,000 candidate signals and five training seeds.
The coefficient analysis in Fig.~2f of the main text uses 2,000 signals.

\textbf{Electrocardiogram (ECG) signals.}
We use the MIT--BIH Arrhythmia Database~\cite{moody2001impact}, excluding
paced records 102, 104, 107, and 217. We select the
modified limb lead II (MLII) and extract $[R-99,R+160)$, where $R$ is the annotated beat
location. Heartbeats are resampled to 256 points without normalization.
We use the DS1/DS2 record partition of de Chazal et al.~\cite{dechazal2004automatic}.
DS1 is partitioned at the record level into training and validation subsets,
with four of its 22 records (108, 114, 207, and 230) reserved for validation.
The same partition is used across all random seeds.
DS2 provides 49,691 heartbeats for testing only. Models use 78 measurements, eight layers, and
256-dimensional codes. Joint learning starts from a model trained with fixed Gaussian sensing
and a Symlet-4 dictionary.

\textbf{DVS-G.}
DVS-Gesture (DVS-G)~\cite{amir2017low} events are integrated using
SpikingJelly~\cite{fang2023spikingjelly} into 16 segments with approximately equal event counts,
producing $16\times2\times128\times128$ tensors. Event counts are transformed using
$\log(1+x)$ and divided by the maximum transformed count in each sample.
All methods use these processed frames as inputs and reconstruction targets.

\textbf{SHD.}
For Spiking Heidelberg Digits (SHD)~\cite{cramer2020heidelberg}, the first 1,000~ms is divided into 25 bins of 40~ms over 700 channels.
Integer event counts are retained without clipping or normalization. Both event datasets use the provided training and test splits. Following previous practice~\cite{Fang_2021_ICCV,baronig2025advancing}, the main results in this paper use the test splits for model selection and evaluation. In Supplementary Note~5, we additionally report results with validation subsets held out from the training data, using the official test sets exclusively for evaluation.

Supplementary Table~\ref{tab:supp_architectures} summarizes the network architectures.
LISTA models for the event datasets have eight layers with weights shared across time steps.
For SHD, the auxiliary and evaluation classifiers use the same temporal convolutional network (TCN) architecture
with independently trained parameters. Group normalization is applied within
the temporal residual blocks, and layer normalization is retained after pooling.

\begin{table*}[htbp]
\centering
\caption{Network architectures. FC, Conv, and TConv denote fully connected,
convolutional, and transposed-convolutional layers; numbers give output widths.
For LISTA, input mappings and update components are listed separately;
connections follow Methods. DVS-G Conv kernels are $3\times3$;
TConv kernels are $4\times4$ with stride 2.
Each residual block contains two spiking layers.
LIF, ReLU, and LSTM denote leaky integrate-and-fire neurons, rectified linear units, and long short-term memory networks, respectively. ANN denotes artificial neural network.}
\label{tab:supp_architectures}
\begingroup
\footnotesize
\setlength{\tabcolsep}{3pt}
\renewcommand{\arraystretch}{1.0}
\begin{tabular}{@{}p{0.14\textwidth}p{0.26\textwidth}p{0.29\textwidth}p{0.25\textwidth}@{}}
\toprule
Module & Synthetic / ECG & DVS-G & SHD \\
\midrule
Sensing & FC$M$ ($M=141/78$)
& Conv16 $\to$ Conv32 $\to$ Conv4; strides 2, 1, 2
& FC256 \\
S-LISTA & Input: FC256; update: FC256 + LIF
& Input: Conv128; update: Conv128 + LIF
& Input: FC700; update: FC700 + LIF \\
ANN LISTA~\cite{gregor2010learning} & Input: FC256; update: FC256 + Soft threshold
& Input: Conv128; update: Conv128 + Soft threshold
& Input: FC700; update: FC700 + Soft threshold \\
ALISTA~\cite{liu2019alista} & Analytical FC256; learned step sizes and Soft threshold
& -- & -- \\
LAMP~\cite{borgerding2017amp} & $L$ FC256 mappings; Soft threshold; Onsager correction & -- & -- \\
Vanilla SNN & --
& Conv64 + ReLU $\to$ Conv128 + LIF; one residual block
& FC256 + LIF $\to$ FC700 + LIF; three residual blocks \\
ANN LSTM & -- & ConvLSTM128 & LSTM512 \\
Readout & Identity (synthetic); FC256 (ECG)
& TConv128 $\to$ TConv64 $\to$ Conv2;
Vanilla SNN: TConv256 $\to$ TConv128 $\to$ Conv2
& FC700 \\
Task classifiers & --
& Conv3D64 $\to$ Conv3D128 $\to$ Conv3D256; global pooling
& Temporal Conv384, dilations 1, 2, 4; group normalization with 32 groups; attention/mean/max pooling \\
\bottomrule
\end{tabular}
\endgroup
\end{table*}

\subsection*{Training and transmission settings}

All models use Adam~\cite{kingma2014adam} with cosine learning-rate annealing.
Synthetic training uses batches of 1,024 signals and two stages of 5,000
and 15,000 updates. The second-stage learning rate is one-quarter of the first-stage rate.
ECG uses batches of 256 and 100-epoch fixed and initial joint-learning runs;
final S-LISTA training uses 200 epochs. DVS-G and SHD use 150 and 100 epochs
with batch sizes of 64 and 512, respectively. Supplementary Table~\ref{tab:supp_learning_rates} gives the joint-training learning rates.
The minimum rate is $10^{-6}$ for ECG and $10^{-7}$ otherwise.

\begin{table*}[htbp]
\centering
\caption{For ECG S-LISTA, the two reconstruction rates correspond to the
measurement-embedding and residual-correction operators, respectively;
for the other datasets, the listed S-LISTA reconstruction rate applies to both. A dash denotes no auxiliary
classifier.}
\label{tab:supp_learning_rates}
\begingroup
\footnotesize
\setlength{\tabcolsep}{4pt}
\renewcommand{\arraystretch}{1.0}
\begin{tabular*}{\textwidth}{@{\extracolsep{\fill}}llcccc@{}}
\toprule
Dataset & Method & Reconstruction & Sensing & Dictionary & Auxiliary classifier \\
\midrule
ECG & ANN LISTA & $3.5\times10^{-4}$ & $10^{-3}$ & $5\times10^{-4}$ & -- \\
& ALISTA & $4\times10^{-3}$ & $2\times10^{-4}$ & $2\times10^{-4}$ & -- \\
& LAMP & $1.5\times10^{-3}$ & $7.5\times10^{-4}$ & $1.5\times10^{-3}$ & -- \\
& S-LISTA & $6\times10^{-4}$ / $1.5\times10^{-4}$
& $4.5\times10^{-4}$ & $9\times10^{-4}$ & -- \\
\midrule
DVS-G & S-LISTA & $3\times10^{-4}$ & $3\times10^{-4}$ & $5\times10^{-4}$ & $3\times10^{-4}$ \\
& Vanilla SNN & $10^{-3}$ & $3\times10^{-4}$ & $5\times10^{-4}$ & $3\times10^{-4}$ \\
& ANN LISTA & $7\times10^{-4}$ & $3\times10^{-4}$ & $5\times10^{-4}$ & $3\times10^{-4}$ \\
& ANN LSTM & $5\times10^{-4}$ & $3\times10^{-4}$ & $5\times10^{-4}$ & $3\times10^{-4}$ \\
\midrule
SHD & S-LISTA & $7.5\times10^{-4}$ & $3\times10^{-4}$ & $10^{-3}$ & $8\times10^{-4}$ \\
& Vanilla SNN & $6\times10^{-4}$ & $3\times10^{-4}$ & $5\times10^{-4}$ & $3\times10^{-4}$ \\
& ANN LISTA / ANN LSTM & $7\times10^{-4}$ & $3\times10^{-4}$ & $5\times10^{-4}$ & $3\times10^{-4}$ \\
\bottomrule
\end{tabular*}
\endgroup
\end{table*}

Synthetic training minimizes the mean per-sample normalized squared error.
Other experiments use mean-squared error with sparse-code regularization
weights of $10^{-4}$ for ECG, $10^{-7}$ for DVS-G, and $10^{-3}$ for SHD.
Task-supervision weights are 0.01 on DVS-G and 3.0 on SHD, with
cross-entropy normalized by the logarithm of the class count.
The event-relaxation temperature decreases exponentially from 1.0 to 0.1.
The ablation without temperature relaxation replaces the temperature-dependent
surrogate gradients with rectangular surrogate gradients at both the sensing
and reconstruction stages, while retaining hard spike generation in the forward pass.
Selected address-event representation (AER) rate weights for S-LISTA and Vanilla SNN are 0.0006 and 0.0014
on DVS-G, and 0.10 and 0.12 on SHD. Full-vector comparisons omit rate
regularization. Reconstruction checkpoints are selected by the lowest
normalized mean squared error (NMSE) on the designated evaluation split. For ECG, checkpoint selection uses only the DS1 validation subset;
DS2 is used only for final evaluation.

With sensing and reconstruction frozen, evaluation classifiers are trained
separately for each method under additive white Gaussian noise (AWGN) at 10~dB: DVS-G uses 100 epochs, batch size 32, and learning rate
$5\times10^{-4}$; SHD uses 150 epochs, batch size 256, and learning rate
$10^{-3}$. Both use weight decay $10^{-4}$ and select the highest-accuracy checkpoint.

Synthetic models use noise-free training and additive measurement noise
at evaluation. ECG uses AWGN at 5~dB for training and uniform 8-bit
quantization over $[-3,3]$; the transmission-budget comparison uses separately trained
2-, 4-, 6-, and 8-bit models, corresponding to 156, 312, 468, and
624 bits per sample, respectively. Event models are trained under AWGN at 10~dB.
Wireless evaluation scans AWGN and Rayleigh fading from $-5$ to 20~dB
with fixed weights and the binary phase-shift keying (BPSK) model in Methods. Main event results
and AER ablations use Rayleigh fading at 20~dB.

AER follows Methods with 16-coordinate blocks. Both DVS-G and SHD transmit
a single binary spike stream with values in $\{0,1\}$, without an additional
polarity bit. ANN baselines use 8-bit measurements.

\subsection*{Evaluation metrics}

NMSE is calculated as
\begin{equation}
\mathrm{NMSE}=10\log_{10}
\frac{\sum_{i=1}^{N_{\mathrm{eval}}}
\|\bar{\hat{\mathbf{x}}}_i-\bar{\mathbf{x}}_i\|_2^2}
{\sum_{i=1}^{N_{\mathrm{eval}}}\|\bar{\mathbf{x}}_i\|_2^2},
\label{eq:supp_nmse}
\end{equation}
where $N_{\mathrm{eval}}$ is the number of evaluation samples,
$\bar{\mathbf{x}}_i=T^{-1}\sum_{t=1}^{T}\mathbf{x}_{i,t}$, and
$\bar{\hat{\mathbf{x}}}_i$ is the corresponding reconstructed average.
We use $T=1$ for synthetic signals and ECG, $T=16$ for DVS-G, and $T=25$
for SHD. Per-sample NMSE uses the same expression with $N_{\mathrm{eval}}=1$ and is reported in dB. Downstream accuracy uses complete
reconstructed sequences. Results for synthetic signals, ECG, and the event datasets are averaged over training seeds 42--46.
Unless otherwise stated, error bars and shaded bands indicate one standard deviation across training seeds.
The ECG waveform reconstructions in Fig.~3a--d are individual examples from seed 42 rather than averages across seeds.

Energy is estimated using 4.6~pJ per multiply--accumulate (MAC) operation and 0.9~pJ per accumulate (AC) operation~\cite{horowitz20141}, as detailed in Methods.
An activation of integer magnitude $n$ contributes $n$ ACs per connection.
Counts cover all time steps and are averaged over samples.
For DVS-G and SHD, estimates use models trained with seed 42 and evaluated under AWGN at 10~dB.
Synthetic and ECG estimates use noise-free measurements and AWGN at 5~dB, respectively, and are averaged over training seeds 42--46.

\section*{Supplementary Note 3: Theoretical analysis of S-LISTA}
\label{supp:theoretical_analysis}

We analyse S-LISTA with its parameters fixed after training. The network
updates and notation follow Supplementary Note 1.

\subsection*{3.1 Inverse problem formulation and assumptions}
\label{supp:inverse_problem}

For the recovery-error analysis, assume that the source signal
$\mathbf{x}_t\in\mathbb{R}^{N}$ at time $t$ admits a sparse representation
\begin{equation}
    \mathbf{x}_t=\boldsymbol{\Psi}\mathbf{z}_t^\star,
    \label{eq:supp_sparse_representation}
\end{equation}
where $\boldsymbol{\Psi}\in\mathbb{R}^{N\times N_z}$ is the dictionary
and $\mathbf{z}_t^\star\in\mathbb{R}^{N_z}$ is the true sparse code.
The dimensions of the source signal and sparse code are denoted by $N$
and $N_z$, respectively. The sparsity of the true code is
$s_t=\|\mathbf{z}_t^\star\|_0$, where $\|\cdot\|_0$ counts nonzero
coordinates.

We model the received measurement vector as
\begin{equation}
    \widetilde{\mathbf{s}}_t^{\rm rx}
    =\mathbf{A}\mathbf{z}_t^\star+\mathbf{n}_t,
    \label{eq:supp_inverse_model}
\end{equation}
where $M$ is the measurement dimension,
$\mathbf{A}\in\mathbb{R}^{M\times N_z}$ maps the sparse code to ideal
measurements, and $\mathbf{n}_t\in\mathbb{R}^{M}$ is the total noise
at the input of S-LISTA.
For continuous linear sensing,
$\mathbf{A}=\boldsymbol{\Phi}\boldsymbol{\Psi}$, where
$\boldsymbol{\Phi}\in\mathbb{R}^{M\times N}$ is the sensing matrix.
In the synthetic experiments, $N_z=N$, $\boldsymbol{\Psi}=\mathbf{I}$,
and $\mathbf{A}$ is the Gaussian measurement matrix.

For spike transmission, $\mathbf{A}$ specifies a linear reference for
the measurement process. Using the transmitted measurement vector
$\mathbf{s}_t^{\rm tx}$ defined in Methods, the total noise can be
written as
\begin{equation}
    \mathbf{n}_t
    =\underbrace{\widetilde{\mathbf{s}}_t^{\rm rx}
      -\mathbf{s}_t^{\rm tx}}_{\text{transmission noise}}
    +\underbrace{\mathbf{s}_t^{\rm tx}
      -\mathbf{A}\mathbf{z}_t^\star}_{\text{encoding noise}},
    \label{eq:supp_effective_noise}
\end{equation}
where the transmission noise is the difference between the received and
transmitted measurement vectors. AWGN and fading affect its distribution
through the bit decisions. The encoding noise measures the deviation of
spike measurements from the linear reference, including the effect of
transmitter memory. The noise bounds used in the recovery analysis must
cover both terms. We do not require $\mathbf{n}_t$ to be Gaussian or
independent of the source.

The inverse problem is to recover $\mathbf{z}_t^\star$ and hence
$\mathbf{x}_t$ from $\widetilde{\mathbf{s}}_t^{\rm rx}$. S-LISTA produces
$\mathbf{z}_t$ through the updates in Supplementary Note 1 and reconstructs
$\widehat{\mathbf{x}}_t=\boldsymbol{\Psi}\mathbf{z}_t$.
The recovery bounds below concern
$\|\mathbf{z}_t-\mathbf{z}_t^\star\|_2^2$ for the resulting network.
No spectral-norm constraint is imposed on the learned operators
$\mathbf{P}_{\rm s}$ and $\mathbf{G}_{\rm s}^{(\ell)}$.

For streaming inputs, each layer retains its previous-frame membrane
state through $\rho_{\rm rx}\mathbf{u}_{t-1}^{(\ell)}$ in
Eq.~\eqref{eq:supp_membrane}, while the accumulated code starts from
$\mathbf{z}_t^{(0)}=\mathbf{0}$ at each frame. Framewise recovery bounds
condition on the membrane states determined by previous received frames. For static inputs, $T=1$ and
$\mathbf{u}_0^{(\ell)}=\mathbf{0}$, so the initial memory term vanishes.

\subsection*{3.2 Boundedness of membrane states and reconstruction outputs}
\label{supp:boundedness}

For a vector, the norm below is the largest absolute value among its
entries. For a matrix, it is the largest sum of absolute entries in a row:
\[
    \|\mathbf{v}\|_\infty=\max_i|v_i|,
    \qquad
    \|\mathbf{W}\|_\infty=\max_i\sum_j|W_{ij}|.
\]
These norms satisfy
$\|\mathbf{W}\mathbf{v}\|_\infty
\leq\|\mathbf{W}\|_\infty\|\mathbf{v}\|_\infty$.

\begin{proposition}[Boundedness of S-LISTA]
\label{prop:supp_boundedness}
Consider S-LISTA in Supplementary Note 1 with fixed finite parameters,
a fixed depth $L\geq2$, and $\mathbf{u}_0^{(\ell)}=\mathbf{0}$.
Assume $0\leq\rho_{\rm rx}<1$, $\vartheta_{\rm rx}^{(\ell)}>0$ for
$\ell=1,\ldots,L$, and $\lambda_{\rm out}\geq0$.
Suppose there is a finite constant $B_{\rm in}$, independent of $t$,
such that $\|\widetilde{\mathbf{s}}_t^{\rm rx}\|_\infty
\leq B_{\rm in}$ for every time step $t\geq1$.
Then all membrane states and reconstruction outputs are bounded by
constants independent of $t$, and for every $t\geq1$,
\begin{equation}
    \|\mathbf{z}_t\|_\infty
    \leq L+
    \frac{
        \|\mathbf{P}_{\rm s}\|_\infty B_{\rm in}
        +(L-1)\|\mathbf{G}_{\rm s}^{(L-1)}\|_\infty
    }{
        \vartheta_{\rm rx}^{(L)}(1-\rho_{\rm rx})
    }.
    \label{eq:supp_output_uniform_bound}
\end{equation}
\end{proposition}

\begin{proof}
Each layer changes each coefficient of the sparse code by at most one
in magnitude.
Since the accumulated code starts from zero at each frame,
Eq.~\eqref{eq:supp_code_accumulation} gives
\begin{equation}
    \mathbf{z}_t^{(\ell)}
    =\sum_{k=1}^{\ell}\boldsymbol{\xi}_t^{(k)},
    \qquad
    \|\mathbf{z}_t^{(\ell)}\|_\infty\leq\ell.
    \label{eq:supp_accumulated_code_bound}
\end{equation}
Using the residual correction in
Eq.~\eqref{eq:supp_residual_correction}, we bound the two terms in
the input current separately. For $\ell=2,\ldots,L$,
\begin{align}
    \|\mathbf{I}_t^{(\ell)}\|_\infty
    &=\left\|
        \mathbf{P}_{\rm s}\widetilde{\mathbf{s}}_t^{\rm rx}
        -\mathbf{G}_{\rm s}^{(\ell-1)}\mathbf{z}_t^{(\ell-1)}
      \right\|_\infty \notag\\
    &\leq
        \|\mathbf{P}_{\rm s}\|_\infty
        \|\widetilde{\mathbf{s}}_t^{\rm rx}\|_\infty
        +\|\mathbf{G}_{\rm s}^{(\ell-1)}\|_\infty
         \|\mathbf{z}_t^{(\ell-1)}\|_\infty \notag\\
    &\leq
        \|\mathbf{P}_{\rm s}\|_\infty B_{\rm in}
        +(\ell-1)\|\mathbf{G}_{\rm s}^{(\ell-1)}\|_\infty.
    \label{eq:supp_layer_current_bound}
\end{align}
For the first layer, Eq.~\eqref{eq:supp_embedding} gives
$\|\mathbf{I}_t^{(1)}\|_\infty
\leq\|\mathbf{P}_{\rm s}\|_\infty B_{\rm in}$.

If no spike is emitted, the reset leaves the membrane potential
unchanged. Otherwise, the spike has the sign of the pre-reset potential,
whose magnitude is at least the threshold. The reset then subtracts
one threshold from this magnitude. Thus, for each neuron $i$,
\begin{equation}
    |u_{t,i}^{(\ell)}|
    =|\bar{u}_{t,i}^{(\ell)}|
      -\vartheta_{\rm rx}^{(\ell)}|\xi_{t,i}^{(\ell)}|
    \leq|\bar{u}_{t,i}^{(\ell)}|.
    \label{eq:supp_reset_magnitude}
\end{equation}
The reset therefore cannot increase the membrane magnitude.
Using Eq.~\eqref{eq:supp_membrane}, we obtain
\begin{equation}
    \|\mathbf{u}_t^{(\ell)}\|_\infty
    \leq\rho_{\rm rx}\|\mathbf{u}_{t-1}^{(\ell)}\|_\infty
        +\|\mathbf{I}_t^{(\ell)}\|_\infty.
    \label{eq:supp_membrane_recursion}
\end{equation}
Applying this inequality recursively from the zero initial state
gives a sum of input currents weighted by their decay factors. Using
Eq.~\eqref{eq:supp_layer_current_bound}, we obtain, for $\ell\geq2$,
\begin{align}
    \|\mathbf{u}_t^{(\ell)}\|_\infty
    &\leq\sum_{t'=1}^{t}
        \rho_{\rm rx}^{\,t-t'}\|\mathbf{I}_{t'}^{(\ell)}\|_\infty
        \notag\\
    &\leq\frac{1-\rho_{\rm rx}^{\,t}}{1-\rho_{\rm rx}}
        \left[
            \|\mathbf{P}_{\rm s}\|_\infty B_{\rm in}
            +(\ell-1)\|\mathbf{G}_{\rm s}^{(\ell-1)}\|_\infty
        \right] \notag\\
    &\leq\frac{
            \|\mathbf{P}_{\rm s}\|_\infty B_{\rm in}
            +(\ell-1)\|\mathbf{G}_{\rm s}^{(\ell-1)}\|_\infty
        }{1-\rho_{\rm rx}},
    \label{eq:supp_membrane_uniform_bound}
\end{align}
where $t'$ indexes time steps from $1$ to $t$. The second inequality uses
the geometric sum
$1+\rho_{\rm rx}+\cdots+\rho_{\rm rx}^{t-1}
=(1-\rho_{\rm rx}^{t})/(1-\rho_{\rm rx})$.
Since $0\leq\rho_{\rm rx}<1$, this sum is at most
$1/(1-\rho_{\rm rx})$, giving the last inequality.
For $\ell=1$, the same argument gives the bound
$\|\mathbf{P}_{\rm s}\|_\infty B_{\rm in}/(1-\rho_{\rm rx})$.
Equation~\eqref{eq:supp_membrane} expresses the pre-reset state as
the decayed previous state plus the current. Both terms are bounded,
so the pre-reset states are also bounded.

Finally, the support mask only retains or removes coefficients, and
the output soft threshold does not increase their magnitudes:
$|\mathcal{S}_{\lambda_{\rm out}}(a)|\leq|a|$ for
$\lambda_{\rm out}\geq0$. Applying these properties to
Eq.~\eqref{eq:supp_readout} gives
\begin{align}
    \|\mathbf{z}_t\|_\infty
    &\leq\left\|
        \mathbf{z}_t^{(L)}+\mathbf{m}_t\odot
        \frac{\mathbf{u}_t^{(L)}}{\vartheta_{\rm rx}^{(L)}}
    \right\|_\infty \notag\\
    &\leq L+
        \frac{\|\mathbf{u}_t^{(L)}\|_\infty}
             {\vartheta_{\rm rx}^{(L)}}.
    \label{eq:supp_readout_magnitude}
\end{align}
Substituting Eq.~\eqref{eq:supp_membrane_uniform_bound} with $\ell=L$
proves Eq.~\eqref{eq:supp_output_uniform_bound}. The reconstructed
signal is bounded because the dictionary is fixed and finite:
\begin{equation}
    \|\widehat{\mathbf{x}}_t\|_\infty
    \leq\|\boldsymbol{\Psi}\|_\infty\|\mathbf{z}_t\|_\infty.
    \label{eq:supp_signal_magnitude}
\end{equation}
\end{proof}

\subsection*{3.3 Recovery error bounds of S-LISTA}
\label{supp:recovery_bounds}

We first bound the recovery error using the spikes emitted by the
trained network. We then consider how sufficiently small input noise
changes the output. Throughout this section, $\|\mathbf{a}\|_2$ denotes
the Euclidean norm, and $\|\mathbf{W}\|_2$ denotes the induced matrix 2-norm.

At step $t$, let $S_t$ be the given support of the true sparse
code, with $s_t$ nonzero coefficients. Assume that their magnitudes
are at most $B$, and that the input noise has norm at most
$\varepsilon$. In the bound calculation, $\mathbf{z}$ denotes a variable
sparse code,
while $\mathbf{z}_t^\star$ denotes the true sparse code. We allow
$\mathbf{z}\in\mathbb{R}^{N_z}$ and noise
$\mathbf{n}\in\mathbb{R}^{M}$ to vary subject to
\begin{equation}
    z_i=0\quad(i\notin S_t),\qquad
    |z_i|\leq B\quad(i\in S_t),\qquad
    \|\mathbf{n}\|_2\leq\varepsilon,
    \label{eq:supp_recovery_prior}
\end{equation}
where $B>0$ and $\varepsilon\geq0$ are specified bounds. The calculation
uses the support and these bounds, but not the individual nonzero
values of $\mathbf{z}_t^\star$. The spikes
$\boldsymbol{\xi}_t^{(\ell)}$, accumulated codes
$\mathbf{z}_t^{(\ell)}$, mask $\mathbf{m}_t$, and previous-step
states $\mathbf{u}_{t-1}^{(\ell)}$ below are fixed at their values
in the network run being analysed.

The emitted spikes also restrict the possible input current.
For any pair $(\mathbf{z},\mathbf{n})$ satisfying
Eq.~\eqref{eq:supp_recovery_prior}, substitute
$\mathbf{A}\mathbf{z}+\mathbf{n}$ into the membrane equation:
\begin{equation}
    \bar{\mathbf{u}}_t^{(\ell)}(\mathbf{z},\mathbf{n})
    =\rho_{\rm rx}\mathbf{u}_{t-1}^{(\ell)}
     +\mathbf{P}_{\rm s}(\mathbf{A}\mathbf{z}+\mathbf{n})
     -\mathbf{G}_{\rm s}^{(\ell-1)}\mathbf{z}_t^{(\ell-1)}.
    \label{eq:supp_bound_membrane}
\end{equation}
The last term is omitted for $\ell=1$.
Equation~\eqref{eq:supp_signed_spike} implies the following necessary
conditions for every layer $\ell$ and neuron $i$:
\begin{equation}
    \begin{cases}
    \bar u_{t,i}^{(\ell)}(\mathbf{z},\mathbf{n})
        \geq\vartheta_{\rm rx}^{(\ell)},
        &\xi_{t,i}^{(\ell)}=+1,\\
    -\vartheta_{\rm rx}^{(\ell)}
        \leq\bar u_{t,i}^{(\ell)}(\mathbf{z},\mathbf{n})
        \leq\vartheta_{\rm rx}^{(\ell)},
        &\xi_{t,i}^{(\ell)}=0,\\
    \bar u_{t,i}^{(\ell)}(\mathbf{z},\mathbf{n})
        \leq-\vartheta_{\rm rx}^{(\ell)},
        &\xi_{t,i}^{(\ell)}=-1.
    \end{cases}
    \label{eq:supp_spike_constraints}
\end{equation}
For zero spikes, including the threshold endpoints enlarges the
allowed range and therefore preserves an upper bound. All these
conditions are linear in $\mathbf{z}$ and $\mathbf{n}$. They are
imposed together with Eq.~\eqref{eq:supp_recovery_prior}; the received
measurement is not additionally fixed while these variables vary.

Before the final soft threshold, define the error expression
\begin{equation}
    \mathbf{e}_t(\mathbf{z},\mathbf{n})
    =\mathbf{m}_t\odot\left[
        \mathbf{z}_t^{(L-1)}
        +\frac{
            \rho_{\rm rx}\mathbf{u}_{t-1}^{(L)}
            +\mathbf{P}_{\rm s}(\mathbf{A}\mathbf{z}+\mathbf{n})
            -\mathbf{G}_{\rm s}^{(L-1)}\mathbf{z}_t^{(L-1)}
        }{\vartheta_{\rm rx}^{(L)}}
    \right]-\mathbf{z}.
    \label{eq:supp_affine_recovery_error}
\end{equation}
The proof below derives this expression from the continuous readout.
Its components are linear functions plus constants, with coefficients
specified by the trained network and $\mathbf{A}$.

To bound the squared error, express $\mathbf{e}_t$ in an orthonormal
basis $\mathbf{Q}=[\mathbf{q}_1,\ldots,\mathbf{q}_{N_z}]$, with
\mbox{$\mathbf{Q}^{\mathsf T}\mathbf{Q}=\mathbf{I}$}:
\begin{equation}
    \mathbf{e}_t(\mathbf{z},\mathbf{n})
    =\sum_{i=1}^{N_z}
        \widetilde e_{t,i}(\mathbf{z},\mathbf{n})\mathbf{q}_i.
    \label{eq:supp_error_basis_expansion}
\end{equation}
Projecting onto $\mathbf{q}_i$ gives
\begin{equation}
    \widetilde e_{t,i}(\mathbf{z},\mathbf{n})
    =\mathbf{q}_i^{\mathsf T}\mathbf{e}_t(\mathbf{z},\mathbf{n})
    =\sum_{j=1}^{N_z}Q_{ji}e_{t,j}(\mathbf{z},\mathbf{n}),
    \label{eq:supp_error_components}
\end{equation}
where $Q_{ji}$ is the entry in row $j$, column $i$ of $\mathbf{Q}$,
and $e_{t,j}$ is the $j$th component of the error in
Eq.~\eqref{eq:supp_affine_recovery_error}. The basis is fixed during
the optimization. The transformed components remain linear functions
plus constants, and their squares sum to the original squared error:
\begin{equation}
    \sum_{i=1}^{N_z}\widetilde e_{t,i}^2
    =\mathbf{e}_t^{\mathsf T}\mathbf{Q}\mathbf{Q}^{\mathsf T}\mathbf{e}_t
    =\mathbf{e}_t^{\mathsf T}\mathbf{e}_t
    =\|\mathbf{e}_t\|_2^2.
    \label{eq:supp_error_norm_preservation}
\end{equation}
In each minimum and maximum below, the same constraints
\eqref{eq:supp_recovery_prior} and
\eqref{eq:supp_spike_constraints} apply:
\begin{equation}
    \underline e_{t,i}
        =\min_{\mathbf{z},\mathbf{n}}
            \widetilde e_{t,i}(\mathbf{z},\mathbf{n}),
    \qquad
    \overline e_{t,i}
        =\max_{\mathbf{z},\mathbf{n}}
            \widetilde e_{t,i}(\mathbf{z},\mathbf{n}).
    \label{eq:supp_error_intervals}
\end{equation}
These are linear optimization problems when $\varepsilon=0$.
For $\varepsilon>0$, they retain linear objectives and include the
Euclidean noise constraint in Eq.~\eqref{eq:supp_recovery_prior}.
Define
\begin{equation}
    U_t
    =\max_{\mathbf{z},\mathbf{n}}
      \sum_{i=1}^{N_z}\Bigl[
        (\underline e_{t,i}
            +\overline e_{t,i})
        \widetilde e_{t,i}(\mathbf{z},\mathbf{n})
        -\underline e_{t,i}
            \overline e_{t,i}
      \Bigr].
      \label{eq:supp_joint_upper_bound}
\end{equation}
The maximization uses one common pair $(\mathbf{z},\mathbf{n})$
for all terms, rather than adding their separate maxima. Its
objective is again linear.

\begin{theorem}[Recovery error bound of S-LISTA]
\label{thm:supp_recovery_bound}
Consider S-LISTA with fixed finite parameters, $L\geq2$,
$\vartheta_{\rm rx}^{(\ell)}>0$, and $\lambda_{\rm out}\geq0$.
If $(\mathbf{z}_t^\star,\mathbf{n}_t)$ satisfies
Eq.~\eqref{eq:supp_recovery_prior}, then the bound constructed from
its network run in Eqs.~\eqref{eq:supp_bound_membrane}--\eqref{eq:supp_joint_upper_bound}
satisfies
\begin{equation}
    \|\mathbf{z}_t-\mathbf{z}_t^\star\|_2^2
    \leq\left(\sqrt{U_t}
        +\lambda_{\rm out}\sqrt{s_t}\right)^2.
    \label{eq:supp_final_recovery_bound}
\end{equation}
\end{theorem}

\begin{proof}
The actual pair $(\mathbf{z}_t^\star,\mathbf{n}_t)$ satisfies
Eq.~\eqref{eq:supp_spike_constraints} by the firing rule.
The optimization constraints therefore admit this pair. They are
closed and bounded by Eq.~\eqref{eq:supp_recovery_prior}, so the
minima and maxima above exist and are finite.

Let $\mathbf{z}_t^{\rm c}$ denote the continuous readout before the
final soft threshold. The mask in Eq.~\eqref{eq:supp_support_mask}
satisfies $\mathbf{m}_t\odot\mathbf{z}_t^{(L)}
=\mathbf{z}_t^{(L)}$. Substituting the accumulation and reset
equations into Eq.~\eqref{eq:supp_readout} gives
\begin{align}
    \mathbf{z}_t^{\rm c}
    &=\mathbf{z}_t^{(L)}+
        \mathbf{m}_t\odot
        \frac{\mathbf{u}_t^{(L)}}{\vartheta_{\rm rx}^{(L)}}
        \notag\\
    &=\mathbf{m}_t\odot\left[
        \mathbf{z}_t^{(L-1)}+\boldsymbol{\xi}_t^{(L)}
        +\frac{\bar{\mathbf{u}}_t^{(L)}
           -\vartheta_{\rm rx}^{(L)}\boldsymbol{\xi}_t^{(L)}}
           {\vartheta_{\rm rx}^{(L)}}
      \right]\notag\\
    &=\mathbf{m}_t\odot\left[
        \mathbf{z}_t^{(L-1)}
        +\frac{\bar{\mathbf{u}}_t^{(L)}}
              {\vartheta_{\rm rx}^{(L)}}
      \right].
      \label{eq:supp_readout_cancellation}
\end{align}
Using Eq.~\eqref{eq:supp_bound_membrane} at the actual pair yields
\begin{equation}
    \mathbf{z}_t^{\rm c}-\mathbf{z}_t^\star
    =\mathbf{e}_t(\mathbf{z}_t^\star,\mathbf{n}_t).
    \label{eq:supp_actual_affine_error}
\end{equation}

We use the linear upper bound for a convex function on an interval,
applied to $x^2$~\cite[Section~3.1.1]{boyd2004convex}.
For any real numbers $a\leq x\leq b$,
$(x-a)(b-x)\geq0$. Expanding this product gives
\begin{equation}
    x^2\leq(a+b)x-ab.
    \label{eq:supp_scalar_square_bound}
\end{equation}
Apply this inequality to each component in
Eq.~\eqref{eq:supp_error_components}, using its bounds from
Eq.~\eqref{eq:supp_error_intervals}. Orthogonality preserves the
sum of squared components, so every allowed pair satisfies
\begin{align}
    \|\mathbf{e}_t(\mathbf{z},\mathbf{n})\|_2^2
    &=\sum_{i=1}^{N_z}
        \widetilde e_{t,i}(\mathbf{z},\mathbf{n})^2\notag\\
    &\leq\sum_{i=1}^{N_z}\Bigl[
        (\underline e_{t,i}
            +\overline e_{t,i})
        \widetilde e_{t,i}(\mathbf{z},\mathbf{n})
        -\underline e_{t,i}
            \overline e_{t,i}
      \Bigr]\notag\\
    &\leq U_t.
    \label{eq:supp_joint_bound_proof}
\end{align}
Evaluating this inequality at the actual pair gives
$\|\mathbf{z}_t^{\rm c}-\mathbf{z}_t^\star\|_2\leq\sqrt{U_t}$.

For $\lambda\geq0$, scalar soft thresholding satisfies
$|\mathcal{S}_{\lambda}(a)-\mathcal{S}_{\lambda}(b)|
\leq|a-b|$~\cite[Sections~2.3 and~6.5.2]{parikh2014proximal}. Indeed, it is continuous and piecewise linear with
slopes zero or one. Applying this property to each coefficient shows
that vector soft thresholding cannot increase Euclidean distance.
Also, it changes each nonzero coefficient of
$\mathbf{z}_t^\star$ by at most $\lambda_{\rm out}$ and leaves
its zero coefficients unchanged. Hence
\begin{align}
    \|\mathbf{z}_t-\mathbf{z}_t^\star\|_2
    &\leq
      \|\mathcal{S}_{\lambda_{\rm out}}(\mathbf{z}_t^{\rm c})
          -\mathcal{S}_{\lambda_{\rm out}}(\mathbf{z}_t^\star)\|_2
      +\|\mathcal{S}_{\lambda_{\rm out}}(\mathbf{z}_t^\star)
          -\mathbf{z}_t^\star\|_2\notag\\
    &\leq\|\mathbf{z}_t^{\rm c}-\mathbf{z}_t^\star\|_2
          +\lambda_{\rm out}\sqrt{s_t}\notag\\
    &\leq\sqrt{U_t}+\lambda_{\rm out}\sqrt{s_t}.
    \label{eq:supp_soft_threshold_error}
\end{align}
Squaring both sides proves Eq.~\eqref{eq:supp_final_recovery_bound}.
\end{proof}

\renewcommand{\figurename}{Supplementary Fig.}
\begin{figure}[H]
    \centering
    \includegraphics[width=0.9\textwidth]{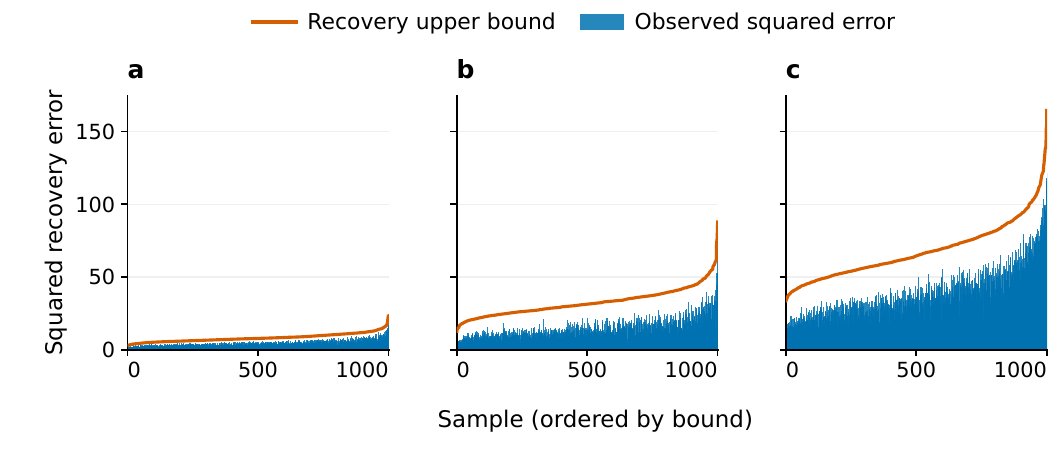}
    \caption{\textbf{Recovery error bounds of S-LISTA on synthetic sparse signals.}
    Results are shown for $L=10$ layers and sparsities
    \textbf{a}, $s=28$; \textbf{b}, $s=42$; and \textbf{c}, $s=56$.
    Each panel contains 1,000 test samples ordered by increasing upper
    bound. The orange curve shows the calculated upper bound in
    Eq.~\eqref{eq:supp_final_recovery_bound}, and the blue bars show
    the observed squared recovery error
    $\|\mathbf{z}_t-\mathbf{z}_t^\star\|_2^2$ in the same sample order.}
    \label{fig:supp_recovery_bound}
\end{figure}

Supplementary Fig.~\ref{fig:supp_recovery_bound} compares the bound
in Eq.~\eqref{eq:supp_final_recovery_bound} with the observed squared
recovery error for the synthetic sparse signals. The bounds remain
on the same scale as the observed errors across the three sparsity
levels. Although the samples are ordered only by their upper bounds,
the observed errors also increase overall, indicating that the bound
reflects differences in recovery error between samples.

The bound above uses the spikes and previous-step states of the
run being analysed. To obtain an explicit relation to noise
amplitude, we next compare two runs of the same network. Superscripts
$\mathrm{clean}$ and $\mathrm{noise}$ denote inputs
$\mathbf{A}\mathbf{z}_{t'}^\star$ and
$\mathbf{A}\mathbf{z}_{t'}^\star+\mathbf{n}_{t'}$, respectively.
The distance of a clean pre-reset potential to the nearer of its
two firing thresholds determines a sufficient noise range in
which its spike is unchanged.

\begin{corollary}[Sensitivity of S-LISTA to small input noise]
\label{cor:supp_small_noise}
Consider these two runs with identical finite initial membrane
states and fixed parameters, with $0\leq\rho_{\rm rx}<1$,
$\vartheta_{\rm rx}^{(\ell)}>0$, and $\lambda_{\rm out}\geq0$.
Suppose $\|\mathbf{n}_{t'}\|_2\leq\varepsilon$ and, for every
$t'=1,\ldots,t$, layer $\ell$, and neuron $i$,
\begin{equation}
    \varepsilon\,
    \frac{1-\rho_{\rm rx}^{\,t'}}{1-\rho_{\rm rx}}
    \sqrt{\sum_{j=1}^{M}(\mathbf{P}_{\rm s})_{ij}^{\,2}}
    <\left|
        \left|\bar u_{t',i}^{(\ell),\mathrm{clean}}\right|
        -\vartheta_{\rm rx}^{(\ell)}
      \right|.
    \label{eq:supp_noise_margin_condition}
\end{equation}
Then their spikes coincide through step $t$, and
\begin{equation}
    \|\mathbf{z}_t^{\mathrm{noise}}
       -\mathbf{z}_t^{\mathrm{clean}}\|_2
    \leq
    \frac{\|\mathbf{P}_{\rm s}\|_2}{\vartheta_{\rm rx}^{(L)}}
    \frac{1-\rho_{\rm rx}^{\,t}}{1-\rho_{\rm rx}}
    \,\varepsilon.
    \label{eq:supp_small_noise_bound}
\end{equation}
\end{corollary}

\begin{proof}
Define the membrane-state difference between the two runs by
\begin{equation}
    \Delta\mathbf{u}_{t'}^{(\ell)}
    =\mathbf{u}_{t'}^{(\ell),\mathrm{noise}}
     -\mathbf{u}_{t'}^{(\ell),\mathrm{clean}}.
    \label{eq:supp_noise_difference_definition}
\end{equation}
Use the same definition for the pre-reset potentials
$\Delta\bar{\mathbf{u}}_{t'}^{(\ell)}$.
We proceed in time order and, within each step, in layer order.
The initial differences are zero.
At a given step $t'$ and layer $\ell$, suppose that all spikes
processed earlier are identical in the two runs. Their accumulated
codes before this layer are then identical. Subtracting their
membrane equations therefore cancels the residual current and gives
\begin{equation}
    \Delta\bar{\mathbf{u}}_{t'}^{(\ell)}
    =\rho_{\rm rx}\Delta\mathbf{u}_{t'-1}^{(\ell)}
      +\mathbf{P}_{\rm s}\mathbf{n}_{t'}
    =\sum_{k=1}^{t'}\rho_{\rm rx}^{\,t'-k}
      \mathbf{P}_{\rm s}\mathbf{n}_k,
    \label{eq:supp_noise_membrane_difference}
\end{equation}
where $k$ indexes steps $1,\ldots,t'$. The second equality uses
the identical reset terms at steps before $t'$. By the Cauchy--Schwarz inequality,
\begin{align}
    |\Delta\bar u_{t',i}^{(\ell)}|
    &\leq\sum_{k=1}^{t'}\rho_{\rm rx}^{\,t'-k}
        |(\mathbf{P}_{\rm s}\mathbf{n}_k)_i|\notag\\
    &\leq\varepsilon\,
        \frac{1-\rho_{\rm rx}^{\,t'}}{1-\rho_{\rm rx}}
        \sqrt{\sum_{j=1}^{M}(\mathbf{P}_{\rm s})_{ij}^{\,2}}.
    \label{eq:supp_noise_neuron_bound}
\end{align}
Condition~\eqref{eq:supp_noise_margin_condition} makes this change
smaller than the distance to either firing threshold. The current
spike is therefore unchanged. The two soft resets subtract the
same quantity, so
$\Delta\mathbf{u}_{t'}^{(\ell)}
=\Delta\bar{\mathbf{u}}_{t'}^{(\ell)}$.
This proves the induction step, starting from the first time step
and the first layer.
Thus all spikes, accumulated codes, and masks coincide through
step $t$.

Using their common mask $\mathbf{m}_t$ and final accumulated code,
the two continuous readouts satisfy
\begin{equation}
    \mathbf{z}_t^{\rm c,noise}-\mathbf{z}_t^{\rm c,clean}
    =\frac{\mathbf{m}_t}{\vartheta_{\rm rx}^{(L)}}\odot
      \sum_{t'=1}^{t}\rho_{\rm rx}^{\,t-t'}
      \mathbf{P}_{\rm s}\mathbf{n}_{t'}.
    \label{eq:supp_noise_readout_difference}
\end{equation}
Masking cannot increase the Euclidean norm, and the final soft
threshold cannot increase the distance between these readouts.
Consequently,
\begin{align}
    \|\mathbf{z}_t^{\mathrm{noise}}
       -\mathbf{z}_t^{\mathrm{clean}}\|_2
    &\leq\frac{1}{\vartheta_{\rm rx}^{(L)}}
        \sum_{t'=1}^{t}\rho_{\rm rx}^{\,t-t'}
        \|\mathbf{P}_{\rm s}\mathbf{n}_{t'}\|_2\notag\\
    &\leq\frac{\|\mathbf{P}_{\rm s}\|_2}
                  {\vartheta_{\rm rx}^{(L)}}
        \frac{1-\rho_{\rm rx}^{\,t}}{1-\rho_{\rm rx}}
        \,\varepsilon.
\end{align}
The bound is linear in the noise limit within the range specified
by Eq.~\eqref{eq:supp_noise_margin_condition}. For a single step,
the geometric factor equals one; across steps, it is at most
$1/(1-\rho_{\rm rx})$. By the triangle inequality, the same
right-hand side also bounds the possible increase in recovery-error
norm relative to the clean run.
\end{proof}

\subsection*{3.4 Sparsity analysis}
\label{supp:sparsity_connection}

We relate the number of nonzero coefficients in the true sparse code
to the firing rate of S-LISTA to examine how signal sparsity affects
spike activity at the receiver.
Let $S_t$ be the support of $\mathbf{z}_t^\star$, containing
$s_t$ nonzero coefficients, and define the firing rate at step $t$ as
\begin{equation}
    r_t=\frac{1}{2N_zL}
        \sum_{\ell=1}^{L}\sum_{i=1}^{N_z}
        |\xi_{t,i}^{(\ell)}|,
    \label{eq:supp_receiver_firing_rate}
\end{equation}
where $|\xi_{t,i}^{(\ell)}|$ counts whether a spike is emitted.
The denominator normalizes the count over $N_z$ coefficients,
$L$ layers, and two spike pathways. Each coefficient emits
at most one spike per layer, so $0\leq r_t\leq1/2$ under this
normalization.

\begin{theorem}[Firing-rate bound with signal sparsity]
\label{thm:supp_sparsity_firing}
Consider the S-LISTA accumulation rule in
Eq.~\eqref{eq:supp_code_accumulation}, with $L\geq2$,
$\mathbf{z}_t^{(0)}=\mathbf{0}$, and
$\xi_{t,i}^{(\ell)}\in\{-1,0,1\}$. Then
\begin{equation}
    r_t\leq\frac{s_t}{2N_z}
    +\frac{
        2\displaystyle\sum_{\ell=1}^{L-1}\sum_{i\notin S_t}
            \left(z_{t,i}^{(\ell)}\right)^2
        +\displaystyle\sum_{i\notin S_t}
            \left(z_{t,i}^{(L)}\right)^2
    }{2N_zL}.
    \label{eq:supp_sparsity_firing_bound}
\end{equation}
\end{theorem}

\begin{proof}
Separate the spike count into positions inside and outside the
true support. There are $s_t$ positions inside $S_t$, and each
can emit at most one spike in each of the $L$ layers. Therefore,
\begin{equation}
    \sum_{\ell=1}^{L}\sum_{i\in S_t}
        |\xi_{t,i}^{(\ell)}|\leq Ls_t.
    \label{eq:supp_support_spike_count}
\end{equation}

For positions outside $S_t$, use the accumulation rule to write
\begin{equation}
    \xi_{t,i}^{(\ell)}
    =z_{t,i}^{(\ell)}-z_{t,i}^{(\ell-1)}.
    \label{eq:supp_spike_accumulation_difference}
\end{equation}
The accumulated coefficients are integers because they start from
zero and each update adds $-1$, $0$, or $+1$. Every integer $a$
satisfies $|a|\leq a^2$. Consequently,
\begin{align}
    |\xi_{t,i}^{(\ell)}|
    &=\left|z_{t,i}^{(\ell)}-z_{t,i}^{(\ell-1)}\right|\notag\\
    &\leq |z_{t,i}^{(\ell)}|+|z_{t,i}^{(\ell-1)}|\notag\\
    &\leq\left(z_{t,i}^{(\ell)}\right)^2
          +\left(z_{t,i}^{(\ell-1)}\right)^2.
    \label{eq:supp_spike_square_bound}
\end{align}
Summing over layers counts every intermediate accumulated value
twice, while the initial and final values appear once. Since
$z_{t,i}^{(0)}=0$, this gives
\begin{align}
    \sum_{\ell=1}^{L}\sum_{i\notin S_t}
        |\xi_{t,i}^{(\ell)}|
    &\leq\sum_{i\notin S_t}\sum_{\ell=1}^{L}
        \left[
            \left(z_{t,i}^{(\ell)}\right)^2
            +\left(z_{t,i}^{(\ell-1)}\right)^2
        \right]\notag\\
    &=2\sum_{\ell=1}^{L-1}\sum_{i\notin S_t}
        \left(z_{t,i}^{(\ell)}\right)^2
      +\sum_{i\notin S_t}\left(z_{t,i}^{(L)}\right)^2.
    \label{eq:supp_outside_support_spike_count}
\end{align}
Adding Eqs.~\eqref{eq:supp_support_spike_count} and
\eqref{eq:supp_outside_support_spike_count} bounds the total spike
count. Dividing by $2N_zL$ proves
Eq.~\eqref{eq:supp_sparsity_firing_bound}.
\end{proof}

Equation~\eqref{eq:supp_sparsity_firing_bound} provides a quantitative
link between signal sparsity and computational sparsity.
The first term depends directly on the number of nonzero coefficients
in the true sparse code. The second term accounts for accumulated
activity outside its support. Noise and membrane states retained
from earlier steps can affect this activity, but the inequality
holds for the resulting spikes without constraints on the trained
weights.

\begin{corollary}[Ideal firing-rate upper bound]
\label{cor:supp_ideal_support_firing}
Under the conditions of Theorem~\ref{thm:supp_sparsity_firing},
if $z_{t,i}^{(\ell)}=0$ for every $i\notin S_t$ and
$\ell=1,\ldots,L$, then
\begin{equation}
    r_t\leq\frac{s_t}{2N_z}.
    \label{eq:supp_ideal_support_rate}
\end{equation}
\end{corollary}

\begin{proof}
All terms outside $S_t$ in
Eq.~\eqref{eq:supp_sparsity_firing_bound} are zero under the stated
condition. The remaining term gives
Eq.~\eqref{eq:supp_ideal_support_rate}.
\end{proof}

When accumulation remains within the true support, the corollary
bounds the firing rate by a quantity proportional to the fraction
$s_t/N_z$ of nonzero coefficients. In the general case,
the additional term in Eq.~\eqref{eq:supp_sparsity_firing_bound}
quantifies how activity outside the true support increases the bound.
This term includes intermediate layers because positive and negative
spikes can cancel in the final accumulated code while still requiring
computation. A zero final coefficient alone therefore does not imply
that its position was inactive in earlier layers. Together with the
recovery-error bound in Theorem~\ref{thm:supp_recovery_bound}, these
results describe both recovery quality and the sparsity of the
receiver's spike activity.

\renewcommand{\figurename}{Supplementary Fig.}
\begin{figure}[H]
    \centering
    \includegraphics[width=0.9\textwidth]{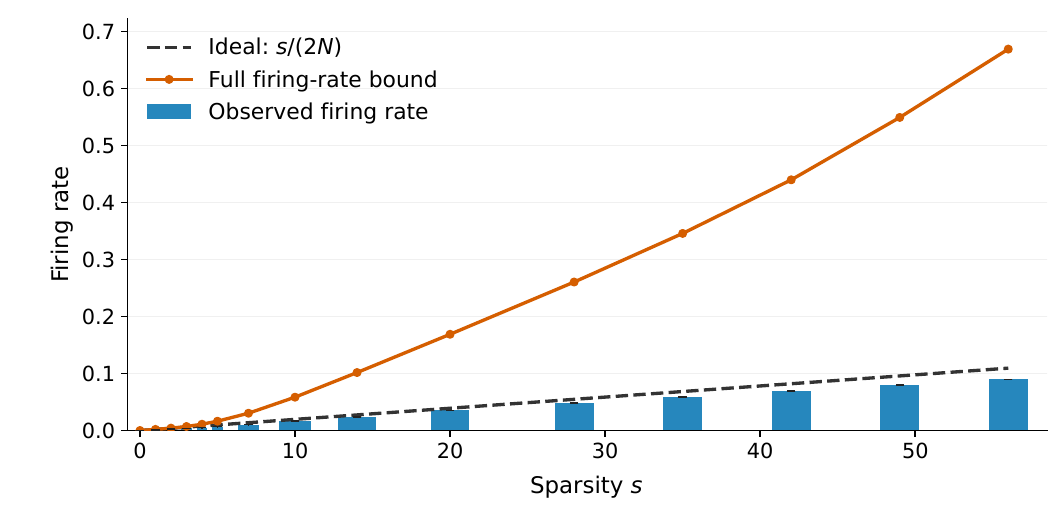}
    \caption{\textbf{Signal sparsity and S-LISTA firing rate.}
    A fixed trained model with $L=10$ layers is evaluated using
    5,000 test samples at each signal sparsity $s$.
    For these synthetic signals, $N_z=N=256$.
    Blue bars show the mean observed firing rate, with error bars
    indicating one standard error of the mean. The dashed line shows
    the ideal bound $s/(2N)$ from
    Eq.~\eqref{eq:supp_ideal_support_rate}. The orange curve shows
    the mean full bound from Eq.~\eqref{eq:supp_sparsity_firing_bound},
    including accumulated activity outside the true support.
    The ideal bound assumes that accumulation remains within the
    true support at every layer.}
    \label{fig:supp_sparsity_firing}
\end{figure}

Supplementary Fig.~\ref{fig:supp_sparsity_firing} compares the mean
firing rate and the bounds using 5,000 test samples at each sparsity
level. With the trained model fixed, the mean firing rate increases
with $s$ and does not exceed the ideal line $s/(2N)$ at any tested
sparsity. These observations provide empirical support for the
connection between signal sparsity and computational sparsity
within the tested setting.

\section*{Supplementary Note 4: Comparison with spiking LCA}

We compare S-LISTA with the generalized spiking locally competitive algorithm (S-LCA)~\cite{du2025generalized} implemented using leaky integrate-and-fire neurons. S-LCA performs sparse recovery through recurrent spiking dynamics and estimates coefficients from accumulated firing rates. Signed coefficients are represented by positive and negative neuron populations using the augmented sensing matrix $[A,-A]$. We use synthetic signals with $N=256$, $M=141$, and $s=28$, following the signal generation procedure in the main text. The regularization parameter is fixed at $\lambda=0.1$ throughout inference, with an integration time step of $0.01$. S-LCA is evaluated over $100$ to $100{,}000$ simulation steps, while S-LISTA uses $20$ unfolded layers and one time step. Both methods are evaluated on the same five independently generated signals, and S-LISTA results are averaged over five training seeds. Inference time is measured on an NVIDIA H100 NVL GPU using FP32 arithmetic and a batch size of one. Synaptic energy is estimated using the same operation costs as in the main text.

\begin{figure}[htbp]
    \centering
    \includegraphics[width=\textwidth]{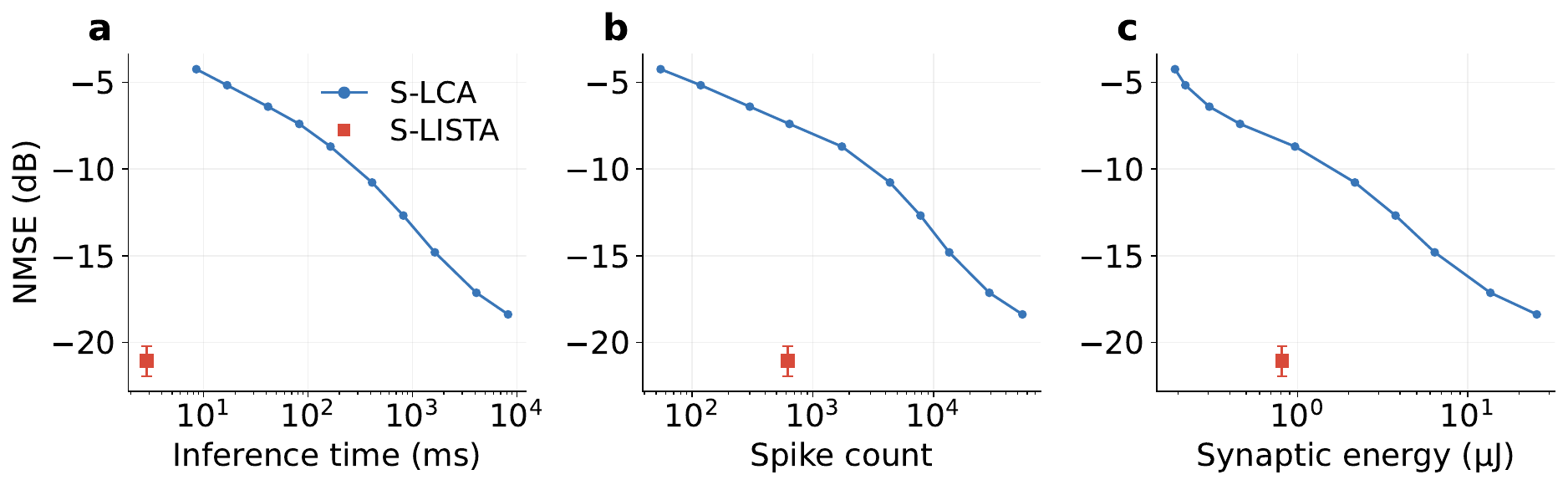}
    \caption{\textbf{Comparison between S-LISTA and S-LCA on synthetic sparse signals.}
    NMSE versus \textbf{a}, GPU inference time; \textbf{b}, spike count per sample; and \textbf{c}, estimated synaptic energy per sample.
    Both methods are evaluated on the same five signals with $N=256$, $M=141$, and $s=28$.
    S-LISTA uses 20 unfolded layers and one time step; error bars indicate one standard deviation across five training seeds.
    S-LCA uses LIF neurons with fixed $\lambda=0.1$, with successive points corresponding to increasing simulation steps from $100$ to $100{,}000$.}
    \label{fig:supp_slca}
\end{figure}

As shown in Supplementary Fig.~\ref{fig:supp_slca}, S-LCA progressively reduces reconstruction error as the simulation duration increases, at the cost of additional execution time and spike activity. Increasing the number of simulation steps from $50{,}000$ to $100{,}000$ improves its NMSE from $-17.14$ to $-18.38$~dB, while increasing inference time from $4.10$ to $8.24$~s and estimated synaptic energy from $13.54$ to $25.35$~$\mu$J per sample. In comparison, S-LISTA achieves an NMSE of $-21.07$~dB with an inference time of $2.85$~ms and an estimated synaptic energy of $0.812$~$\mu$J per sample. Relative to S-LCA at $100{,}000$ steps, S-LISTA achieves $2.69$~dB lower NMSE while using $96.8\%$ less estimated synaptic energy and reducing the mean spike count from approximately $54{,}765$ to $620$ per sample. These results illustrate how reconstruction based on learned algorithm unfolding can reduce the computational cost associated with recurrent spiking convergence.

\section*{Supplementary Note 5: Evaluation with an independent validation set}

We repeat the main DVS-G and SHD experiments with approximately 20\% of the official training data held out for validation, grouped by source recording for DVS-G and stratified by class for SHD. The remaining data are used for training, and the official test sets are unchanged. All other experimental settings are retained. Reconstruction models and downstream classifiers are selected using validation NMSE and accuracy, respectively, under AWGN at 10~dB, and evaluated under Rayleigh fading at 20~dB. Supplementary Table~\ref{tab:dvsg_heldout} shows that S-LISTA maintains comparable classification accuracy to the ANN baselines on DVS-G. On SHD, it achieves the lowest NMSE, with classification accuracy 1.55--1.99 percentage points below the ANN baselines. On both datasets, S-LISTA uses one-eighth of the transmission bits of the ANN baselines.

\begin{table*}[h]
\centering
\caption{\textbf{DVS-G and SHD results with independent validation sets.}
Results use full-vector transmission under Rayleigh fading at 20~dB and are reported as means $\pm$ standard deviations over training seeds 42--46, using a fixed data split for each dataset.}
\label{tab:dvsg_heldout}
\begingroup
\footnotesize
\setlength{\tabcolsep}{5pt}
\renewcommand{\arraystretch}{1.08}
\begin{tabular*}{\textwidth}{@{\extracolsep{\fill}}llccc@{}}
\toprule
Dataset & Method & NMSE (dB) & Accuracy (\%) & Bits/sample \\
\midrule
DVS-G & S-LISTA
& $-12.16 \pm 0.20$
& $\mathbf{94.65 \pm 0.94}$
& $\mathbf{65{,}536}$ \\
DVS-G & Vanilla SNN
& $-9.47 \pm 0.21$
& $89.86 \pm 2.63$
& $\mathbf{65{,}536}$ \\
DVS-G & ANN LISTA
& $-14.69 \pm 0.09$
& $94.03 \pm 1.08$
& $524{,}288$ \\
DVS-G & ANN LSTM
& $\mathbf{-14.78 \pm 0.10}$
& $94.44 \pm 0.89$
& $524{,}288$ \\
\midrule
SHD & S-LISTA
& $\mathbf{-16.65 \pm 0.07}$
& $89.22 \pm 0.67$
& $\mathbf{6{,}400}$ \\
SHD & Vanilla SNN
& $-12.17 \pm 0.23$
& $87.27 \pm 1.22$
& $\mathbf{6{,}400}$ \\
SHD & ANN LISTA
& $-16.32 \pm 0.05$
& $90.77 \pm 0.70$
& $51{,}200$ \\
SHD & ANN LSTM
& $-16.07 \pm 0.07$
& $\mathbf{91.21 \pm 0.45}$
& $51{,}200$ \\
\bottomrule
\end{tabular*}
\endgroup
\end{table*}